\pdfoutput=1
\documentclass[a4paper,fleqn,usenatbib]{mnras}

\usepackage{newtxtext,newtxmath}
\usepackage[T1]{fontenc}
\usepackage{ae,aecompl}

\usepackage{graphicx}	
\usepackage{amsmath}	
\usepackage{amssymb}	
\usepackage{cleveref}

\usepackage[normalem]{ulem} 

\usepackage{subfig}

\title{Herschel-ATLAS : The spatial clustering of low and high redshift submillimetre galaxies}

\author[A. Amvrosiadis et al.]{
A. Amvrosiadis$^{1}$\thanks{E-mail: AmvrosiadisA@cardiff.ac.uk},
E. Valiante$^{1}$,
J. Gonzalez-Nuevo$^{2}$, 
S. J. Maddox$^{1,3}$,
M. Negrello$^{1}$,  \newauthor
S. A. Eales$^{1}$,
L. Dunne$^{1}$,
L. Wang$^{4, 5}$,
E. van Kampen$^{6}$,
G. De Zotti$^{7}$,
M. W. L. Smith$^{1}$, \newauthor
P. Andreani$^{6}$,
J. Greenslade$^{8}$, 
C. Tai-An$^{8}$,
M. J. Micha{\l}owski$^{3,9}$
\\
\\
$^{1}$School of Physics and Astronomy, Cardiff University, The Parade, Cardiff CF24 3AA, UK \\
$^{2}$Departamento de Fìsica, Universidad de Oviedo, C. Federico Garcia Lorca 18, 33007 Oviedo, Spain \\
$^{3}$Institute for Astronomy, The University of Edinburgh, Royal Observatory, Blackford Hill, Edinburgh EH9 3HJ, UK \\
$^{4}$SRON Netherlands Institute for Space Research, Landleven 12, 9747 AD, Groningen, The Netherlands l.wang@sron.nl \\
$^{5}$Kapteyn Astronomical Institute, University of Groningen, Postbus 800, 9700 AV, Groningen, The Netherlands \\
$^{6}$ESO, Karl-Schwarzschild-Str. 2, D-85748 Garching bei Muenchen, Germany \\
$^{7}$INAF?Osservatorio Astronomico di Padova, Vicolo Osservatorio 5, I-35122 Padova, Italy \\
$^{8}$Astrophysics Group, Imperial College, Blackett Laboratory, Prince Consort Road, London SW7 2AZ, UK \\
$^{9}$Astronomical Observatory Institute, Faculty of Physics, Adam Mickiewicz University, ul.~S{\l}oneczna 36, 60-286 Pozna{\'n}, Poland \\
}

\pubyear{2017}

\begin{document}
\label{firstpage}
\pagerange{\pageref{firstpage}--\pageref{lastpage}}
\maketitle

\begin{abstract}

We present measurements of the angular correlation function of sub-millimeter (sub-mm) galaxies (SMGs) identified in four out of the five fields of the \emph{Herschel Astrophysical Terahertz Large Area Survey} \emph{(H-ATLAS)} $-$ \emph{GAMA-9h}, \emph{GAMA-12h}, \emph{GAMA-15h} and \emph{NGP} $-$ with flux densities $S_{250 \mu m}$$>$$30$ mJy at 250 $\mu$m. We show that galaxies selected at this wavelength trace the underlying matter distribution differently at low and high redshifts. We study the evolution of the clustering finding that at low redshifts sub-mm galaxies exhibit clustering strengths of $r_0\sim 2-3$ $h^{-1}$Mpc, below $z<0.3$. At high redshifts, on the other hand, we find that sub-mm galaxies are more strongly clustered with correlation lengths $r_0=8.1\pm0.5$, $8.8\pm0.8$ and $13.9\pm3.9$ $h^{-1}$Mpc at $z=1-2$, $2-3$ and $3-5$, respectively. We show that sub-mm galaxies across the redshift range $1<z<5$,  typically reside in dark-matter halos of mass of the order of $\sim$ $10^{12.5}-10^{13.0} h^{-1}M_{\odot}$ and are consistent with being the progenitors of local massive elliptical galaxies that we see in the local Universe.

\end{abstract}


\section{Introduction}  \label{sec:section_1}

Sub-millimetre galaxies \citep[SMGs, e.g.][and references therein]{Smail1997, Blain2002} are considered to be among the most intensively star-forming objects in the Universe. The ultraviolet (UV) radiation from their newly born stars is absorbed by the dust and then re-emmited at far-infrared (FIR) / sub-millimetre (sub-mm) wavelengths. The total infrared luminosities $(L_{\textit{IR}})$ of some of the brightest SMGs can reach values higher than a few times $10^{12} L_{\odot}$ and sometimes higher than $10^{13} L_{\odot}$ (comparable to Ultra-Luminous InfraRed Galaxies (ULIRGs; Lonsdale et al. 2006) we see in the local Universe) with inferred star formation rates of $\sim$$1000 \, M_{\odot}/$yr \citep[e.g.][]{Swinbank2014}. The shape of their spectral energy distribution (SED) at these wavelengths (Rayleigh-Jeans limit) approximates a power-law that decreases with increasing wavelength, meaning that it is subject to a strong negative K-corrections \citep[e.g.][]{Casey2014}. As a result, these objects are predominantly found at high redshift, in the range $z\sim 1-3$ \citep{Chapman2005, Coppin2006, Simpson2014, Chen2016a}, although a substantial evolution in the co-moving number density of ULIRGS between $z = 0$ and $z = 2-3$ has also been reported (Daddi et al. 2005).

Despite the great success in characterising their properties \citep[see][ and references therein]{Casey2014}, the evolutionary stages and the nature of these high redshift SMGs still remain largely unknown. Various galaxy evolution models have been proposed to explain the morphological transformation and quenching of these objects. The most prevailing scenarios are merger-driven galaxy evolution models, which follow the evolution of both the disc and the spheroidal components of galaxies \citep{Almeida2011}, models where the star formation is fuelled by steady accretion of large amounts of cold gas \citep{Dave2010} and a self-regulated galaxy evolution model \citep{Granato2004,Lapi2011,Cai2013}. 

The evolution of a galaxy population can be constrained from the measurement of its clustering strength, which provides information on the the masses of dark matter halos that these galaxies reside in. There have been numerous clustering studies of SMG's identified in the short \citep[250-500 $\mu$m;][]{Cooray2010, Maddox2010, Mitchell-Wynne2012, vanKampen2012} and long \citep[850-1100 $\mu$m;][]{Webb2003, Blain2004, Scott2006, Weiss2009, Williams2011, Hickox2012, Chen2016a, Chen2016b, Wilkinson2017} submillimetre bands. Similar information can be extracted from the clustering of the unresolved FIR/sub-mm galaxies, through the measurement of the angular power spectrum of CIB anisotropies \citep[][]{Viero2009,Amblard2011,Viero2013,PlanckCollaboration2014_XXX}.

The most accurate determination of the clustering properties of SMG's up to date has been performed by \cite{Chen2016b}. The authors used a sample of $\sim$3000 SMGs with redshifts in the range $z\sim 1-5$, which were selected using a color selection technique \citep{Chen2016a}, Optical-Infrared Triple Color (OIRTC), to preferentially select faint SMG's ($S_{850}<2\,$mJy) in the K-band from the the UKIRT Infrared Deep Sky Survey \citep[UKIDSS;][]{Lawrence2007} Ultra Deep Survey (UDS). In their study they concluded that SMG's, selected with the OIRTC technique, are strongly clustered residing in halos with typical halo masses of the order of $M_h \sim10^{13} h^{-1}M_{\odot}$ across the probed redshift range. However, these sources were not individually detected in the sub-mm wavebands and the evidence that these galaxies are SMGs was based on observations with Atacama Large Millimeter/submillimeter Array (ALMA) training set (a subset of the objects predicted to be brighter sub-mm sources), which implied that the OIRTC method is 87\% efficient.

More recently, \cite{Wilkinson2017} studied the clustering properties of SMG's which were identified using the 850 $\mu$m maps of the UDS field from the SCUBA-2 Cosmology Legacy Survey \citep[S2CLS;][]{Geach2013}. The authors used a sample of 610 SMGs for which they found counterparts using a combination of radio imaging and the optical/infrared selection technique of \cite{Chen2016a}. Using ALMA observations of the brightest sources, they estimate an 80\% successful SMG counterpart identification. However, due to the sparse number density of SMGs the authors relied on a cross-correlation technique, with a more abundant K-band selected sample, to measure their clustering properties. Their analysis yield similar results to \cite{Chen2016b} for $z>2$ SMGs, in terms of the halo masses that these galaxies reside to, but for SMGs found in the redshift range $1<z<2$ they reported a downsizing effect where the SMG activity is shifted to halos with typical halo masses of the order of $M_h \sim10^{12} h^{-1}M_{\odot}$. 

In addition, both \cite{Chen2016b} and \cite{Wilkinson2017} performed the clustering analysis for typical star-forming and passive galaxies, identified in the same field using their colors.  This is important in order to place the clustering results of SMGs in the broader context of galaxy evolution. However, both these studies were unable to significantly differentiate SMG clustering properties from more typical star-forming galaxies identified in the same redshift range. In addition, \cite{Hickox2012} using a sample of 126 SMGs selected at 870$\mu$m from the \emph{Large APEX Bolometer Camera (LABOCA)} sub-mm survey of the \emph{Extended Chandra Deep Field-South (ECDFS)} concluded that the clustering properties of high redshift SMGs are consistent with measurements for optically selected quasi-stellar objects (QSOs). Their findings support evolutionary scenarios in which powerful starburst and QSOs occur in the same systems. In all these studies, high redshift SMGs reside in dark matter halos of the order of $\sim$$10^{13} h^{-1}M_{\odot}$ and seem to be consistent with being the progenitors of massive elliptical galaxies that we see in the local universe.

In order to improve the already existing measurements of the angular clustering signal of SMG we need much larger survey areas to increase the number of detected sources and to obtain accurate redshift information. Concerning the first requirement, the \emph{Herschel Astrophysical Terahertz Large Area Survey (H-ATLAS)} (which cover an area of more than $\sim$600 deg$^2$) provides almost 3 orders of magnitude improvement in covered area compared to surveys conducted at 850$\mu$m \citep{Chen2016a,Chen2016b,Wilkinson2017}. As for the later requirement, the challenge is to identify optical/near-infrared (NIR) counterparts to the sub-mm sources in order to obtain relatively well-constrained photometric or spectroscopic redshifts. This is especially challenging due to the low angular resolution at sub-mm wavelengths, which results in large positional uncertainties for the sub-mm sources. We thoroughly discuss in Section~\ref{sec:H-ATLAS_DATA} how we approach this issue.

Nevertheless, these aforementioned studies provide a unique contribution to the field, enabling for the first time the characterisation of the clustering properties of SMGs as a function of redshift and their role in galaxy evolution scenarios. However, they do not provide a complete picture as they fold in biases linked to the selection of SMGs at these particular wavelengths, rendering it essential to conduct similar studies at different sub-mm wavebands. In this paper we will study the clustering properties of SMG's identified at 250$\mu$m in the H-ATLAS survey, with flux densities $S_{250\mu m}>30\,$mJy. Throughout this paper, we assume a flat $\Lambda$CDM cosmological model with the best-fit parameters derived from the Planck Observatory \citep{PlanckCollaboration2016}, which are $\Omega_m$ = 0.307, $H_0$ = 69.3 km s$^{-1}$ Mpc$^{-1}$ and $\sigma_8=0.816$.

\section{H-ATLAS Data} \label{sec:H-ATLAS_DATA}

The \textit{Herschel Astrophysical Terahertz Large Area Survey} \citep[H-ATLAS;][]{Eales2010} is a survey of $\sim\,$660 deg$^2$ in five far-infrared (far-IR) to submillimeter (sub-mm) photometric bands $-$ 100, 160, 250, 350 and 500 $\mu$m $-$ with the Photoconductor Array Camera and Spectrometer \citep[PACS;][]{Poglitsch2010} and Spectral and Photometric Imaging Receiver \citep[SPIRE;][]{Griffin2010} cameras, which was carried out with the Hershel Space Observatory \citep{Pilbratt2010}. The survey is comprised of five different fields, three of which are located on the celestial equator \citep[GAMA fields;][]{Valiante2016} covering in total an area of 161.6 deg$^2$. The other two fields are centred on the North and South Galactic Poles \citep[NGP and SGP fields;][]{Smith2017} covering areas of 180.1 deg$^2$ and 317.6 deg$^2$, respectively.

The H-ATLAS fields were selected to minimise bright continuum emission from dust in the Galaxy, as seen by the \emph{Infrared Astronomical Satellite} (IRAS) at 100 $\mu$m. Complementary multi-wavelength data for these fields are provided by surveys spanning ultraviolet (UV) to mid-infrared (mid-IR) regimes. In particular for the GAMA fields $-$ \emph{GAMA-9h}, \emph{GAMA-12h}, \emph{GAMA-15h} $-$ optical spectroscopic data are provided by the Galaxy and Mass Assembly survey \citep[GAMA;][]{Driver2009}, the Sloan Digital Sky Survey \citep[SDSS;][]{Abazajian2009} and the 2-Degree-Field Galaxy Redshift Survey \citep[2dFGRS;][]{Colless2001}, while optical photometric data are provided by the Galaxy Evolution Explorer \citep[GALEX;][]{Martin2005} and the Kilo-Degree Survey \citep[KiDS;][]{de_Jong2015}. Besides optical imaging and spectroscopy, imaging data at near-infrared (near-IR) wavelengths are available from the UKIRT Infrared Deep Sky Survey Large Area Survey \citep[UKIDSS-LAS;][]{Lawrence2007}, the Wide-field Infrared Survey Explorer \citep[WISE;][] {Wright2010} and the VISTA Kilo-Degree Infrared Galaxy Survey \citep[VIKING;][]{Edge2013}. In addition, radio-imaging data in the fields are provided by the Faint Images of the Radio Sky at Twenty-cm survey and the NRAO Very Large Array Sky Survey. The multi-wavelength coverage of the \emph{NGP} and \emph{SGP} fields is less extensive. The \emph{NGP} field is covered in the optical by the SDSS and in near-IR by the UKIDSS-LAS while the \emph{SGP} field is covered in the optical by KiDS and in the near-IR by VIKING.

The source catalogues of the H-ATLAS fields, which are presented in \cite{Valiante2016} and \cite{Maddox2018} for the GAMA and NGP/SGP fields respectively, are created in three stages. Firstly, the emission from dust in our Galaxy, which is contained in all \emph{Herschel} images, needs to be removed before the source extraction process. We used the \emph{Nebuliser}\footnote{The \emph{Nebuliser} algorithm was developed by the Cambridge Astronomical Survey Unit, which can be found at
http://casu.ast.cam.ac.uk/surveys-projects/software-release/background-filtering} algorithm, in order to remove this emission from the SPIRE images in all the three wavebands (more details can be found in \cite{Valiante2016} for how the algorithm works). Secondly, the Multiband Algorithm for source Detection and eXtraction (MADX; Maddox et al., in prep.) was used to identify 2.5$\sigma$ peaks in the 250 $\mu$m maps and to measure the flux densities at the position of those peaks in all the SPIRE bands. Before the source extraction, however, the maps were filtered with a matched-filter technique \citep{Chapin2011} in order to reduce instrumental and confusion noise. Finally, only sources with a signal-to-noise ratio $\ge$4 in at least one of the three SPIRE bands were kept in the final catalogue. The 4$\sigma$ detection limit at 250 $\mu$m for a point source ranges from 20mJy in the deepest regions of the maps (where tiles overlap) to 36mJy in the non-overlapping regions. 

Having extracted our sub-mm sources from our \emph{Herschel} maps, ideally we would like to find the counterparts of these sources in other wavelengths. Identifying counterparts to these sub-mm sources, however, is a challenging task. Using likelihood ratio techniques \cite{Bourne2016} identified SDSS optical counterparts to the sub-mm sources in the \textit{GAMA} fields at $r < 22.4$ with a 4$\sigma$ detection at 250$\, \mu$m. The quantity $R$ \emph{(reliability)} corresponds to the probability that a potential counterpart is associated with a \emph{Herschel} source. They find reliable counterparts $(R\, \ge \,0.8)$ for 44,835 sources (39 per cent). In addition, \cite{Furlanetto2018} performed the same analysis for the \textit{NGP} field and obtained optical counterparts for 42,429 sources (37.8 per cent).  One potential caveat of this methodology however, is that it gives a artificially higher likelihood of association for high-z sub-mm sources that are gravitationally lensed by local galaxies or large-scale structure \citep{Bourne2014}.

Finally, we removed local extended sources from our final extracted source catalogue. These sources were selected from the \emph{Herschel} catalogues for having a non-zero aperture radius in either of the three \emph{SPIRE} wavebands. Custom aperture photometry was carried out by \cite{Valiante2016} for the \emph{GAMA} fields and \cite{Maddox2018} for the \emph{NGP} and \emph{SGP} fields. The number of local extended sources are 231, 226, 284, 889 and1452 in the \textit{GAMA-9h}, \textit{GAMA-12h}, \textit{GAMA-15h}, \textit{NGP} and \textit{SGP} fields, respectively.\iffalse}\fi

\subsection{Redshift distribution of sub-mm sources}

The redshift distribution of our sources is an essential ingredient in our clustering analysis. It is used to project the angular correlation function, $w(\theta)$, in order to recover the spatial correlation function, $\xi(r)$, from which the clustering properties of our galaxy population are determined.

\begin{table}
	\centering{
	\caption{Herschel-ATLAS sources with measured spectroscopic redshift from CO observations.}
	\begin{tabular}{lcccc} 
	\hline\\[-6.0mm]
  	\hline
	 H-ATLAS ID & $z_{\textit{spec}}$ & $z_{\textit{phot}}$ & Ref. \\
	\hline
	J134429.5$+$303034 &  2.30 & 2.31 & H12 \\
	J114637.9$-$001132 &  3.26 & 2.81 & H12 \\
	J132630.1$+$334408 &  2.95 & 3.89 & H-p \\
	J083051.0$+$013225 &  3.63 & 3.19 & R-p \\
	J125632.5$+$233627 &  3.57 & 3.56 & R-p \\
	J132427.0$+$284450 &  1.68 & 2.32 & G13 \\
	J132859.2$+$292327 &  2.78 & 2.81 & K-p \\
	J084933.4$+$021442 &  2.41 & 2.91 & L-p \\
	J125135.3$+$261458 &  3.68 & 3.63 & K-p \\
	J113526.2$-$014606 & 3.13 & 2.28 & H12 \\
	J133008.6$+$245900 & 3.11 & 2.36 & R-p \\
	J142413.9$+$022303 & 4.28 & 4.24 & C11 \\
	J141351.9$-$000026 & 2.48 & 285 & H12 \\
	J090311.6$+$003907 & 3.04 & 3.54 & F11 \\
	J132504.4$+$311534 & 1.84 & 2.12 & R-p \\
	J133846.5$+$255055 & 2.34 & 2.69 & R-p \\
	J132301.7$+$341649 & 2.19 & 2.58 & R-p \\
	J091840.8$+$023048 & 2.58 & 3.06 & H12 \\
	J133543.0$+$300402 & 2.68 & 2.76 & H-p \\
	J091304.9$-$005344 & 2.63 & 2.73 & N10 \\
	J115820.1$-$013752 & 2.19 & 3.21 & H-p \\
	J113243.0$-$005108 & 2.58 & 3.92 & R-p \\
	J142935.3$-$002836 & 1.03 & 0.56  & P13 \\
	J090740.0$-$004200 & 1.58 & 1.05 & L12 \\
	J085358.9+015537 & 2.09 & 1.84 & P13 \\
	J090302.9-014127 & 2.31 & 1.97 & L12 \\
	\hline
	\hline
	\label{tab:HATLAS_spectroscopic_info}
	\end{tabular}
	\\
	}
	\begin{flushleft} 
	\textbf{Notes:} The last column corresponds to references for the CO spectroscopic redshifts: N10 = \cite{Negrello2010}, C11 = \cite{Cox2011}, F11 = \cite{Frayer2011}, H12 = \cite{Harris2012}, L12 = \cite{Lupu2012}, B13 = \cite{Bussmann2013}, G13 = \cite{George2013}, P13 = \cite{Pearson2013}, H-p = Harris et al. (prep), R-p = Riechers et al. (prep), K-p = Krips et al. (prep), L-p = Lupu et al. (prep).
	\end{flushleft} 
\end{table} 

The standard approach used to estimate photometric redshifts, when only IR to sub-mm photometric data are available (in this case the SPIRE \textit{250, 350 \& 500 $\mu$m} flux densities), is to fit a \emph{calibrated} SED template to each source in our sample. This approach has been adopted in many previous studies \citep[][]{Pearson2013,Ivison2016,Bakx2018}.

The first step to this procedure is to determine the best-fit values of the parameters of the SED template. We adopt as our SED template a modified blackbody spectral energy distribution, consisting of two dust components with different temperatures. This is given by,
\begin{equation}
S_{\nu}=A_{\textit{off}} \left[ B_{\nu}(T_h)\nu^{\beta} + \alpha B_{\nu}(T_c)\nu^{\beta} \right] \, ,
\end{equation}
where $S_{\nu}$ is the flux at the rest-frame frequency $\nu$, $A_{\textit{off}}$ is the normalisation factor, $B_{\nu}$ is the Planck blackbody function, $\beta$ is the dust emissivity index, $T_h$ and $T_c$ are the temperatures of the hot and cold dust components, and $\alpha$ is the ratio of the mass of the cold to hot dust. In order to compute our sub-mm photometric redshifts, $z_{\textit{phot}}$, we use the parameters found by \cite{Pearson2013}: $T_h=46.9$, $T_c=23.9$, $\alpha=30.1$ and $\beta=2$.

\begin{figure}
\includegraphics[width=0.475\textwidth, height=0.265\textheight]{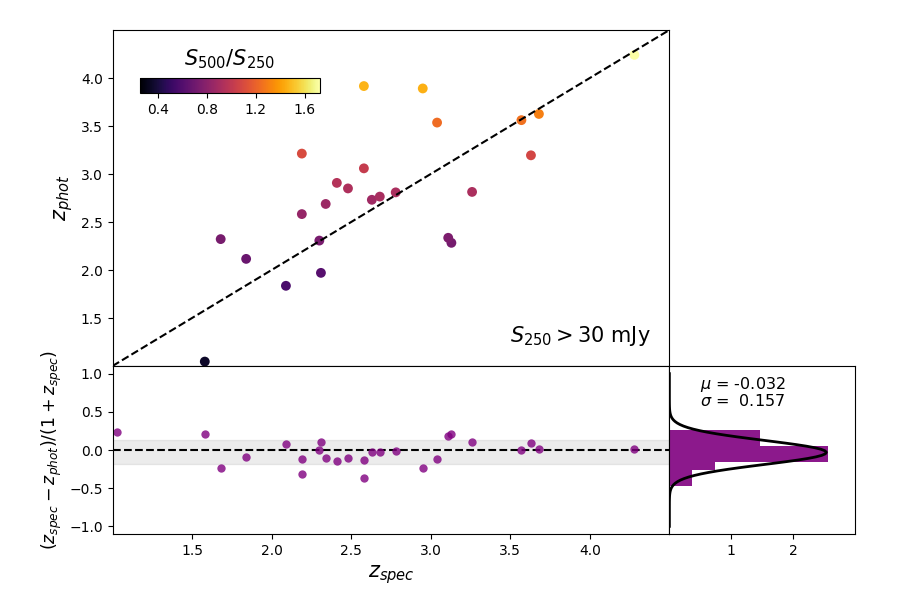} \\
\caption{Scatter plot of $(z_{\textit{spec}} - z_{\textit{phot}})/(1+z_{\textit{spec}})$ against the spectroscopic redshift, $z_{\textit{spec}}$, for sources with CO spectroscopic redshifts in the redshift range, $1<z<5$, which are listed in Table~\ref{tab:HATLAS_spectroscopic_info}. For this comparison we exclude identified QSOs, as it has been shown that the photometric redshift estimation methodology is only reliable for starburst galaxies \citep{Pearson2013}. In the lower right corner we show the histogram of $\Delta z/(1+z)$ values, as well as the mean and standard deviation from fitting a Gaussian distribution (black curve) to this histogram. Finally, in the upper panel we show a scatter plot of $z_{\textit{spec}}$ vs $z_{\textit{phot}}$, where the points are color-coded based on the flux density ratio $S_{500}/S_{250}$ at $500$ and $250$ $\mu$m, respectively. The black dashed line shows the 1:1 relation.}
\label{fig:Zphot_Zspec}
\end{figure}

We evaluate the accuracy of our sub-mm photometric redshift estimates using the same sample of source. In Figure~\ref{fig:Zphot_Zspec} we show $(z_{\textit{spec}} - z_{\textit{phot}})/(1+z_{\textit{spec}})$ vs $z_{\textit{spec}}$, finding that the template performs reasonably well and does not introduce any systematic offset. Fitting a Gaussian distribution to the histogram of $\Delta z/(1+z)$, shown in the lower right corner of Figure~\ref{fig:Zphot_Zspec}, we find a mean of -0.03 with a standard deviation of $\sigma_{\Delta z / (1+z)} = 0.157$ and no outliers. Similar conclusions were drown by Ivison et al. (2016), where the authors used different templates to evaluate their performance. In the top panel of Figure~\ref{fig:Zphot_Zspec} we see that higher redshift sub-mm sources have preferentially redder colors as expected, where the points are color-coded based on the flux density ratio $S_{500}/S_{250}$ at $500$ and $250$ $\mu$m, respectively.

Finally, in order to construct the redshift distributions in Figure 2 we adopted the following procedure.: (i) if  R$\, < \,$0.8 we used the sub-mm photometric redshifts that were determined from our SED fitting methodology. (ii) if R$\, \ge \,$0.8 we further applied an additional cut in redshift quality parameter Q (see Driver et al. 2011 for a detailed definition of the redshift quality parameter Q). If Q$\, \ge \,$3 we used the optical spectroscopic redshift, otherwise we used the optical photometric redshift. In some few cases where R$\, \ge \,$0.8 but none of the above information was available we used sub-mm photometric redshifts.  We need to note here that this selection only concerns the clustering analysis of our low redshift sample since the completeness of our counterpart identification method drops significantly above $z>0.3$ to the point where our high redshift sample $(z>1)$ is completely dominated by sub-mm sources with no counterparts.

\begin{figure*}
\includegraphics[width=0.925\textwidth, height=0.375\textheight]{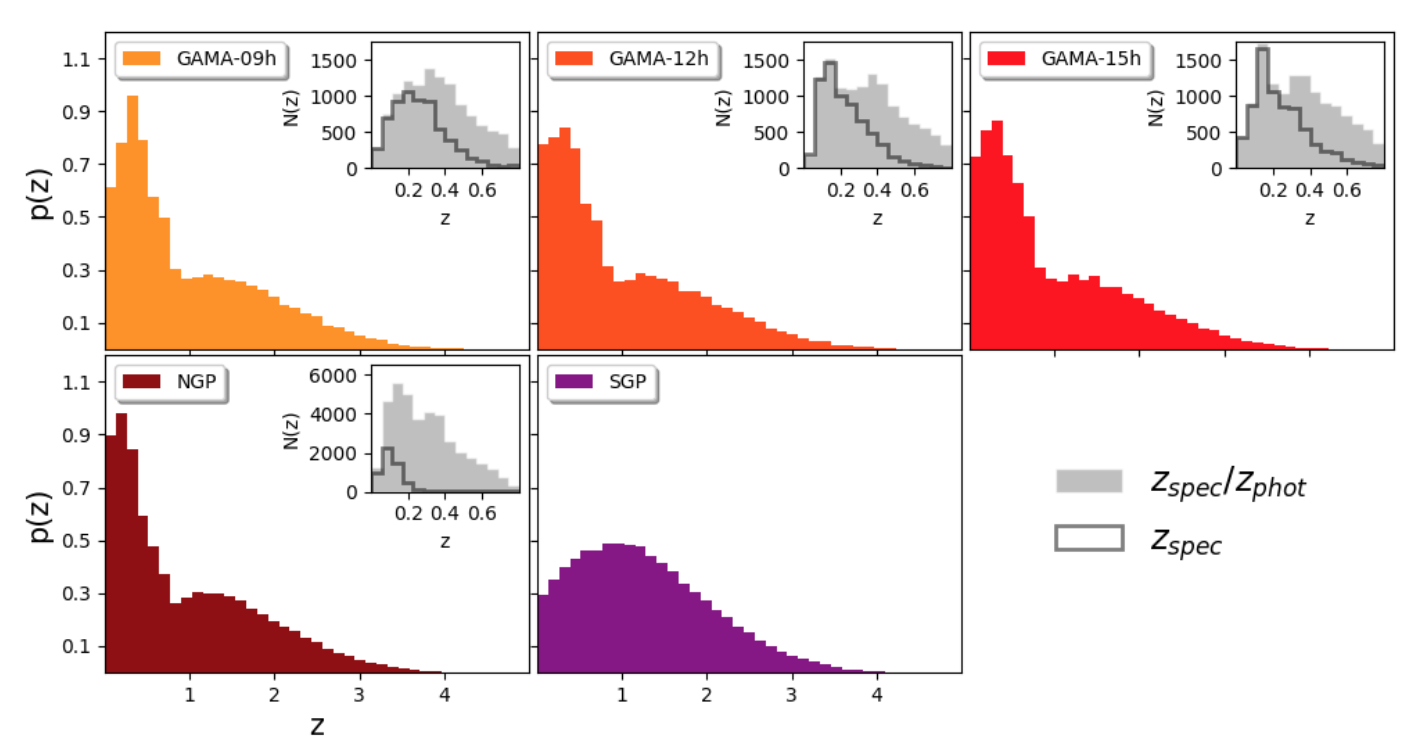} \\
\caption{The redshift distribution of sub-mm sources detected in the five fields of the \emph{H-ATLAS} survey: \emph{GAMA-09h} (top-left), \emph{GAMA-12h} (top-middle), \emph{GAMA-15h} (top-right), \emph{NGP} (bottom-left) and \emph{SGP} (bottom-middle). The histograms are normalised so that the area is equal to unity. The inset plot in each panel shows a zoom into the low redshift range of the redshift distribution of our sub-mm sources with identified counterparts. The grey histogram corresponds to sources with either an optical photometric or spectroscopic redshift (see the main text for more details) while the black histogram corresponds to sources with only optical spectroscopic redshift of quality $Q\ge3$. For the case of the \emph{SGP} field, only sub-mm photometric redshifts are available.}
\label{fig:REDSHIFT_DISTRIBUTION_HATLAS_FIELDS}
\end{figure*}
Figure~\ref{fig:REDSHIFT_DISTRIBUTION_HATLAS_FIELDS} shows the redshift distribution of our sub-mm sources for all H-ATLAS fields, following the procedure outlined above. The inset plot in each panel shows a zoom into the low redshift range of the redshift distribution of our sub-mm sources with identified counterparts. The grey histogram corresponds to sources with either an optical photometric or spectroscopic redshift while the black histogram corresponds to sources with only optical spectroscopic redshift of quality $Q\ge3$. The counterpart identification analysis has not been performed as yet for the SGP field and so in this case we only show the redshift distribution of sub-mm photometric redshifts. One thing to note here is the lack of spectroscopic redshifts in the \emph{NGP} field compared to the \emph{GAMA} fields, which are complemented by the GAMA survey \citep{Driver2009}. 

We can clearly see that our sample of 250$\mu$m selected sub-mm galaxies contains different galaxy populations at low and high redshifts (see \cite{Pearson2013} where the authors performed simulations to show that these are in fact two different galaxy populations rather than being a bias of the sub-mm photometric redshift estimation methodology). On the one hand, the low redshift peak around z$\, \sim \,$0.2 - 0.3 is mostly associated with typical star-forming galaxies (see \cite{Bourne2016} and \cite{Furlanetto2018} for more details on the multi-wavalength properties of H-ATLAS galaxies with identified counterpart), while 15-30\% would be classified as passive galaxies based on their optical colors \citep{Eales2018}. On the other hand, the broader part of the distribution in the redshift range $z>1$ is associated with sub-mm galaxies \citep{Chapman2005}.

\begin{figure}
\includegraphics[width=0.985\columnwidth, height=0.125\textheight]{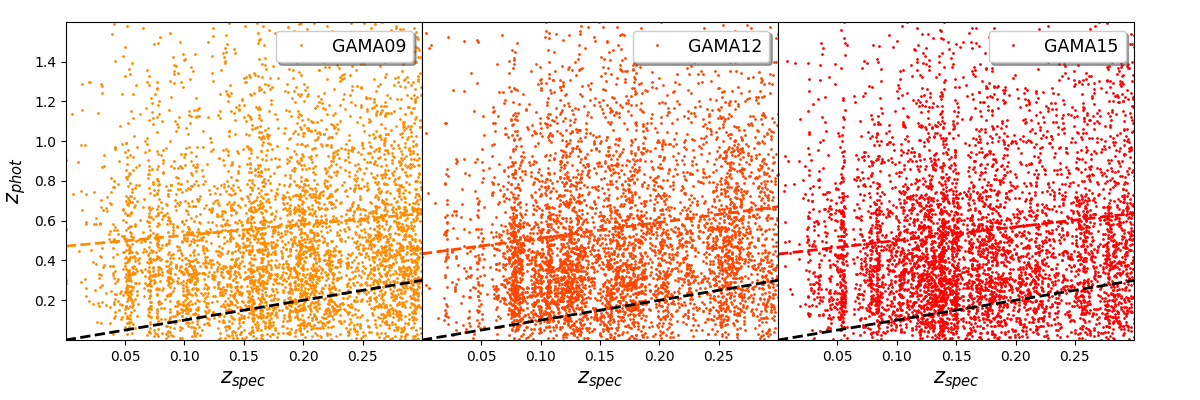} \\
\caption{\iffalse\textbf{\fi Comparison of spectroscopic, $z_{\textit{spec}}$, and photometric, $z_{\textit{phot}}$, redshifts in the redshift range $z<0.3$, for submillimeter sources with identified counterparts. The black dashed lines show the 1:1 relation, while the colored dashed lines show the best-fit line that goes through the data points.\iffalse}\fi }
\label{fig:zspec_vs_zphot_GAMA}
\end{figure}

\subsection{Efftects of sub-mm photometric redshifts}  \label{sec:SubmmPhotometricRedshiftsErrors}

One caveat of using a FIR/sub-mm SED template fitting approach to estimate photometric redshifts for our sub-mm sources, is that the redshifts of low-redshift sources are significantly overestimated. This is clearly demonstrated in Figure~\ref{fig:zspec_vs_zphot_GAMA}, where we show the comparison of $z_{\textit{spec}}$ vs $z_{\textit{phot}}$ for source in the three GAMA field in the redshift range $z<0.3$, with optically identified counterparts. This comparison highlights the importance of identifying the optical counterparts of low-redshifts sub-mm galaxies, when one wishes to measure the clustering properties of high-redshift $(z>1)$ sub-mm galaxies. This is the main reason why we choose not to include the SGP field in the analysis that follows.

In addition, the errors of our sub-mm photometric redshifts, $z_{\textit{phot}}$, which are derived from the SED fitting methodology, are in most cases quite large. This means that when a tomographic analysis of the clustering is performed, a single source can be found in more than one redshift bins. If this effect is not accounted for properly, it can lead to severe biases. \cite{Cowley2017} demonstrated that seemingly similar correlation functions (from observations and simulations of SMGs) result in significantly different clustering properties due to the incorrect normalisation of the correlation function of the underlying dark-matter distribution that these galaxies are tracing.

In order to account for the effect of random errors in photometric redshift estimates on $dN/dz$, following the analysis by Budav{\'a}ri et al. (2003), we estimate the redshift distribution $p(z|W)$ of galaxies selected by our window function $W(z_{\textit{ph}})$, as
\begin{equation}
p(z|W)=p(z) \int dz_{\textit{ph}} W(z_{\textit{ph}})p(z_{\textit{ph}}|z) \,
\end{equation}
where $p(z)$ is the initial redshift distribution, $W(z_{\textit{ph}})$ is a top-hat window function where $W=1$ for $z_{\textit{ph}}$ in the selected redshift interval $z_{\textit{min}}<z<z_{\textit{max}}$ and $W=0$ otherwise, and $p(z_{\textit{ph}}|z)$ is the probability that a source with true redshift $z$ has a photometric redshift $z_{\textit{ph}}$. The function $p(z_{\textit{ph}}|z)$ is parametrised as a Gaussian distribution with a zero mean and variance $(1+z)\sigma_{\Delta z/(1+z)}$,
\begin{equation}
p(z_{\textit{ph}}|z)=\frac{1}{\sqrt{\, 2 \pi (1+z)^2\sigma^2_{\Delta z/(1+z)}}}\exp\left\{-\frac{\left(z-z_{\textit{ph}}\right)^2}{2 (1+z)^2\sigma^2_{\Delta z/(1+z)}}\right\} \,
\end{equation}
where the dispersion is taken to be $\sigma_{\Delta z/(1+z)}=0.15$ as determined from the comparison our sub-mm photometric redshifts and a sample of 26 sources with reliable CO spectroscopic redshifts.

This correction is only relevant for the clustering analysis of our high redshift sample $(z>1)$. This is because our sample is completely dominated by sources with only sub-mm photometric redshift information. Therefore, in this case, the initial redshift distribution, $p(z)$, is estimated by excluding sources with identified counterparts (i.e. those with reliability $R\ge0.8$).
\iffalse}\fi


\section{Clustering Analysis} \label{sec:Clustering_Analysis}

In this section we describe the methodology we followed in order to measure the angular clustering signal.

\subsection{The Angular two-point correlation function} \label{sec:ACF}

The angular two-point auto-correlation function (ACF), $w(\theta)$, is a measure of the excess probability, compared with a random distribution, of finding a galaxy at an angular separation $\theta$ from another, $P(\theta) = N \left[1 + w(\theta)\right]$, where $N$ is the surface density of galaxies. To calculate the angular two-point autocorrelation function we use the \cite{LandySzalay1993} estimator, 
\begin{equation} \label{eq:LandySzalayEstimator}
w(\theta) = \frac{\textit{DD}(\theta) - 2 \, \textit{DR}(\theta) + \textit{RR}(\theta)}{\textit{RR}(\theta)} ,
\end{equation}
where $\textit{DD}(\theta)$ is the number of data-data pairs, $\textit{DR}(\theta)$ is the number of data-random pairs and $\textit{RR}(\theta)$ is the number of random-random pairs, each at separation $\theta$. The $\textit{DR}(\theta)$ and $\textit{RR}(\theta)$ are normalised to have the same number of total pairs as $\textit{DD}(\theta)$, so that given $N_{\textit{D}}$ sample sources and $N_{\textit{R}}$ random points then $\textit{DR}(\theta) = \left[(N_{\textit{D}} - 1)/2N_{\textit{R}}\right]N_{\textit{DR}}(\theta)$ and $\textit{RR}(\theta) = \left[N_{\textit{D}}(N_{\textit{D}} - 1)/N_{\textit{R}}(N_{\textit{R}} - 1)\right]N_{\textit{RR}}(\theta)$, where $N_{\textit{DR}}(\theta)$ and $N_{\textit{RR}}(\theta)$ are the original counts. 

The error on $w(\theta)$ at each angular separation, which is associaciated with the \cite{LandySzalay1993} estimator, is defined as
\begin{equation} \label{eq:LandySzalayEstimator_Errors}
\sigma_w^2 = \frac{\left(1+w\right)^2}{\textit{DD}}.
\end{equation}
However, these errors are considerably underestimated as the variance only accounts for the shot noise from the sample of the random points (which is folded in the measurement of $w$) and the Poisson uncertainties of the DD counts. For a more accurate representation of the errors we consider a 'delete one jackknife' resampling method \citep{Norberg2009}, which also account for systematic uncertainties due to the field-to-field variations. 

In order to implement this approach the area of each field was divided into $N_{\textit{sub}}$ circular sub-regions (as seen in Figure~\ref{fig:NGP_GRID}), each with a radius of $\sim$120 arcmin. Similarly to \cite{GonzalezNuevo2017} we allowed for a 30\% overlap between sub-regions and about less than 10\% of each sub-region did not contain any sources (essentially falling outside of the image). These constraints were introduced in order to maximise the usable area and resulted in 4 independent sub-regions in each of the GAMA fields and 15 for the NGP field (as shown in Figure~\ref{fig:NGP_GRID}). 

Each jackknife sample is defined by discarding, in turn, each of the  $N_{\textit{sub}}$ sub-regions into which each field has been split. The covariance matrix for the $N_{\textit{sub}}$ jackknife resamplings is then estimated using,

\begin{equation} \label{eq:CovarianceMatrix}
C_{i, j} = \frac{N_{\textit{sub}} - 1}{N_{\textit{sub}}} \sum_{k=1}^{N_{\textit{sub}}} \left( w(\theta_i)^k - \bar{w}(\theta_i) \right) \left( w(\theta_j)^k - \bar{w}(\theta_j) \right) \, ,
\end{equation}
where $w(\theta_{i, j})^k$ are the auto-correlation functions measured in each jackknife realisation and $\bar{w}(\theta_{i,j})$ is the average auto-correlation function from all jackknife realisations.

We also corrected the measured correlation function for the integral constraint \citep[IC;][]{RocheEales1999}. Assuming that the true correlation function $w(\theta)$ can be described as a power-law model, $w_{\textit{model}}(\theta) = A\theta^{-\gamma}$,  the observed one will be given by
\begin{equation} \label{eq:IC}
w(\theta) = w_{\textit{model}}(\theta) - \textit{IC} \, ,
\end{equation}
where the IC can be numerically evaluated \citep{Adelberger2005} using the RR counts from,
\begin{equation} \label{eq:IC}
\textit{IC} = \frac{\sum_i \textit{RR}(\theta_i) w_{\textit{model}}(\theta_i)}{\sum_i \textit{RR}(\theta_i)} \, .
\end{equation}
The best-fit values for the power-law model, $w_{\textit{model}}$, from which the IC correction was evaluated, were determined by restricting the angular distance range to $\theta > 4\,$arcmin.\iffalse}\fi

\subsection{Construction of the random catalogues}

\begin{figure*}
\includegraphics[width=0.675\textwidth, height=0.455\textheight]{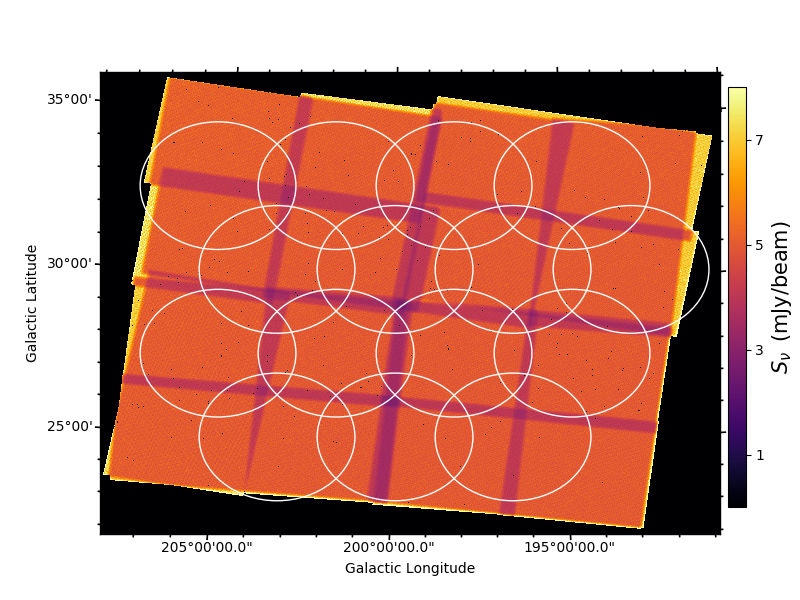} \\
\caption{The filtered variance map of the \emph{NGP} field. The circular areas correspond to the 15 individual sub-regions that the field is divided, in order to perform the "delete one jackknife" resampling method. The black holes in the map indicate the regions covered by extended sources that were masked out.}
\label{fig:NGP_GRID}
\end{figure*}

We mentioned is Section~\ref{sec:H-ATLAS_DATA} that local extended sources were removed from our \emph{Herschel} catalogues, prior to calculating $w(\theta)$. Consequently, we need to account for the removal of these sources when constructing our random catalogues. This is accomplished by masking out the regions covered by extended source in order to avoid placing random sources in those regions. The masked regions were elliptical in the case where a custom aperture was created (using the minor semi-axis as well as the position angle; see section 5.2 in \cite{Valiante2016} for details), otherwise they were circular (see Figure~\ref{fig:NGP_GRID}).

The random catalogues were then created by drawing 10 times more points, than in our real catalogue of sources, from a uniform distribution. 

In practise, however, our noise maps are not completely uniform (as seen in Figure~\ref{fig:NGP_GRID}), due to overlapping scanned regions. It is important that these non-uniformities not be imprinted on the measured clustering signal. We consider a similar approach to that adopted by \cite{Maddox2010}, where we incorporated the noise (instrumental + confusion) information, while making sure we conserve the number counts of our real catalogues. This was achieved as follows: (a) a flux was chosen randomly using the cumulative probability distribution of fluxes of our real sources, (b) a random position was generated on the image, (c) the local noise was estimated as the quadratic sum of the instrumental noise in that pixel and the confusion noise \citep[see Table 3;][ for the GAMA fields]{Valiante2016}, (d) we kept the source if it's flux, perturbed by a Gaussian deviate equal to the total local noise estimate, was greater than 4$\sigma$ otherwise the process was repeated starting from (a)\iffalse}\fi. The measurement of the angular correlation function using random catalogues generated this way, however, shows no significant difference compared to the simple uniform random catalogues. This is due to the fact that we apply a cut in flux-density at 250$\mu$m, which ensures that the fluxes of these sources are not significantly boosted.

\subsection{The real-space correlation length} \label{sec:correlation_length}

The simplest way to interpret the clustering strength of a galaxy population is to estimate its correlation length, $r_0$. We will determine this value for our SMG population at different redshift slices. We assume that the spatial correlation function, $\xi(r)$, is described by a power-law,
\begin{equation}
\xi(r)=\left(\frac{r}{r_0}\right)^{-\gamma}
\end{equation}
where r is the comoving distance between two points, $r_0$ is the correlation length and $\gamma$ is the power-law index.

The angular correlation function, parametrised as power-law model, $w(\theta) = A_w \theta^{-\delta}$, can be deprojected using the Limber approximation \citep{Limber1954} to yield a measurement on the correlation length over different redshift bins. This conversion is performed as follows,
\begin{equation} \label{eq:Real_Space_CorrelationLength}
r_0^{\gamma}(z) = A_{\textit{w}} \left\{ \frac{H_0 H_{\gamma}}{c} \frac{\int_{z_i}^{z_j} N^2(z) (1+z)^{\gamma-(3+\epsilon)} \chi^{1-\gamma}(z) E(z) dz}{\left( \int_{z_i}^{z_j} N(z) dz \right)^2} \right\}^{-1}\, ,
\end{equation}
where the value $\epsilon=\gamma-3$ is assumed, which corresponds to a constant clustering in comoving coordinates. In addition, 
\begin{equation} 
H_{\gamma} = \Gamma\left(\frac{1}{2}\right)\Gamma\left(\frac{\gamma-1}{2}\right) / \Gamma\left(\frac{\gamma}{2}\right)
\end{equation}
with $\Gamma(x)$ being the gamma function and $\chi(z)$ is the radial comoving distance which can be computed from,
\begin{equation}
\chi(z) = \frac{c}{H_0} \int_0^z \frac{dz'}{E(z')} \, .
\end{equation}
where $H_0$ is the Hubble constant and $E^2(z) =  \Omega_{m,0}(1+z)^3 + \Omega_{\Lambda, 0}(1+z)^{3(1+w)}$. Finally, $N(z)$ is the number of sources per unit of redshift interval within a solid angle. The redshift distributions are determined differently for the analysis of our low- and high-redshift samples.

\section{Results} \label{sec:Results}

In this section we present our results of the angular auto-correlation function, $w(\theta)$, for source samples selected with at least a 4$\sigma$ detection at 250-$\mu$m ($\sim$30 mJy). This selection criteria ensures that there are no significant photometry issues with the sources used in our analysis.\iffalse}\fi 
The measurements were performed for evenly spaced logarithmic bins of angular separation in the range $0.5'<\theta<50'$, where the lower limit comes from the FWHM of the SPIRE instrument's PSF at 250$\mu$m ($0.3'$; Griffin et al. 2010).\iffalse}\fi
In addition, as discussed in Section~\ref{sec:H-ATLAS_DATA}, our sample of sources is comprised of different galaxy populations at low and high redshifts. Therefore, we will examine these two cases individually in the sections that follow. 

\subsection{Evolution of Clustering with redshift for $z<0.3$ SMGs} \label{sec:Clustering_evolution_lowZ}

\begin{figure*}
\includegraphics[width=0.975\textwidth, height=0.405\textheight]{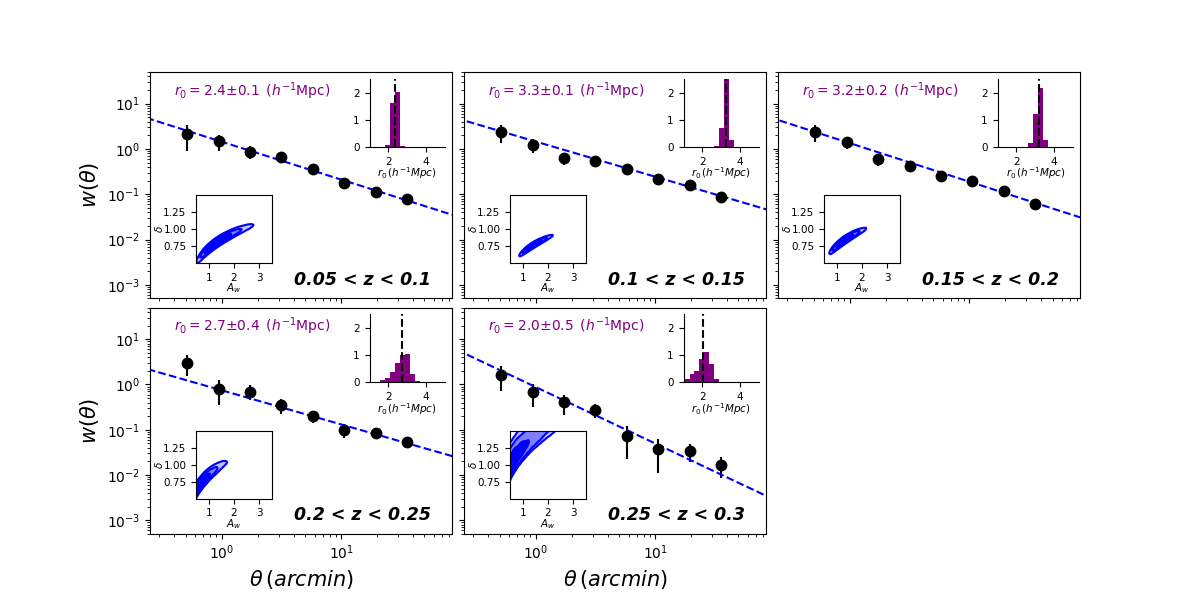} \\
\caption{The angular correlation function of sub-mm galaxies for each redshift slice in the redshift range $z<0.3$. The dashed lines show the best-fit two-parameter model, $w(\theta)=A_{\textit{w}}\theta^{-\delta}$, where the best-fit values can be found in Table~\ref{tab:Angular_Correlation_Function_zDependance_HATLAS_lowZ_Table}. The inset plot in the lower left corner in each panel corresponds to the 1, 1.5 and 2$\sigma$ contours in the fitted $(A_{\textit{w}}, \delta)$ parameter space. The inset plot in each panel shown the histogram of correlation length values which were derived from our bootstrap method. The black dashed vertical line in the inset plot of each panel, indicates the mean of the distribution which is also shown in the upper left corner in each panel. \iffalse {\color{red}There was no correction applied in the measured clustering signal} \fi}
\label{fig:Angular_Correlation_Function_zDependance_HATLAS_lowZ}
\end{figure*}

\begin{table*}
	\centering{
	\caption{Results of Clustering Analysis for $z<0.3$ SMGs}
	\begin{tabular}{lccccc} 
	\hline
  	\hline
	& $0.05 < z < 0.10$ $\,$ & $0.10 < z < 0.15$ $\,$ & $0.15 < z < 0.20$ $\,$ & $0.20 < z < 0.25$ $\,$ & $0.25 < z < 0.30$ \\
	\hline
	$N_{\text{gal}}$ & 6225 & 8284 & 8385 & 7914 & 6744 \\[0.25mm]
	\\
	$A_{\textit{w}}$ & $1.44^{+0.67}_{-0.50}$ & $1.45^{+0.37}_{-0.30}$ & $1.34^{+0.41}_{-0.36}$ & $0.74^{+0.48}_{-0.32}$ & $0.88^{+1.25}_{-0.57}$ \\[0.25mm]
	\\
	$\delta$ & $0.83^{+0.14}_{-0.14}$ & $0.77^{+0.08}_{-0.08}$ & $0.85^{+0.09}_{-0.10}$ & $0.75^{+0.18}_{-0.19}$ & $1.24^{+0.43}_{-0.40}$ \\[0.25mm]
	\\
	$r_0 \, (h^{-1}\textit{Mpc})$ & 2.4$\pm$0.1 & 3.3$\pm$0.1 & 3.2$\pm$0.2 & 2.7$\pm$0.3 & 2.0$\pm$0.5 \\[0.25mm]
	\hline
	\hline
	\label{tab:Angular_Correlation_Function_zDependance_HATLAS_lowZ_Table}
	\end{tabular}
	\\
	}
\end{table*}

The clustering evolution of sub-mm sources selected at 250$\mu$m, in the low redshift regime $(z<0.3)$, has previously been studied by \cite{vanKampen2012}. In their study, the authors used a sample of sources selected from the H-ATLAS Science Demonstration Phase \textit{(SDP)} field at a 5$\sigma$ significance level accounting for both instrumental and confusion noise. This resulted in a flux-density cut of $S>33$ mJy/beam at 250$\mu m$. Additional selection criteria that were introduced in their study, specifically concerning the reliability of counterpart identification and the quality of optical spectroscopic redshifts, were identical to the ones introduced here.\iffalse}\fi

In this section we repeat the analysis of  \cite{vanKampen2012} for a sample of sources selected at 250$\mu m$, from the \emph{GAMA+NGP} fields of the \emph{H-ATLAS} survey. The \emph{SGP} field was not used in this analysis since the optical counterpart identification analysis has not been performed as yet for this field. Similarly to \cite{vanKampen2012} we start our analysis at redshift $z\sim0.05$ where the redshift distribution starts to pick up (see Figure~\ref{fig:REDSHIFT_DISTRIBUTION_HATLAS_FIELDS}) and end at $z\sim0.3$ where the completeness starts to drop sharply \citep[see ][]{Bourne2016}. We use a width size of $\Delta z$$\sim$$0.05$ which results in five individual redshift bins.  

Our clustering measurements are shown in Figure~\ref{fig:Angular_Correlation_Function_zDependance_HATLAS_lowZ}, where each panel corresponds to a different redshift bin indicated at the bottom right corner. The redshift distribution of sources for which this measurement corresponds to, are shown as the grey histograms in each panel of Figure~\ref{fig:REDSHIFT_DISTRIBUTION_HATLAS_FIELDS}. One thing to note is that the clustering signal in the \emph{NGP} field is slightly weaker compared to the \emph{GAMA} fields, which is probably due to the lack of spectroscopic redshifts coverage.\iffalse}\fi

In order to model the clustering signal we used a two-parameter power-law model, $w(\theta) = A_w \theta^{-\delta}$, and performed an MCMC fitting method using the \textit{emcee} package \citep{Foreman-Mackey2013}. The 1, 1.5 and 2$\sigma$ contours of the fitted $(A_w, \delta)$ parameter space are shown in the bottom left inset plot in each panel of Figure~\ref{fig:Angular_Correlation_Function_zDependance_HATLAS_lowZ}. The resulting best-fit values for the parameters of our model in each redshift bin are presented in Table~\ref{tab:Angular_Correlation_Function_zDependance_HATLAS_lowZ_Table}. These correspond to predictions that are shown as blue dashed lines in each panel of Figure~\ref{fig:Angular_Correlation_Function_zDependance_HATLAS_lowZ}. The power-law slopes in all redshift bins are broadly consistent with that of normal star-forming galaxies, $\delta$$\sim$$0.8$ \citep{Zehavi2011}. Although 20-30\% of H-ATLAS galaxies have the red optical colours typical of traditional passive galaxies, \cite{Eales2018} show that these are still star-forming galaxies, although with a significant old stellar population. Therefore, it is not surprising that we find a clustering signal typical of star-forming galaxies.\iffalse}\fi

The clustering length, $r_0$, in each redshift slice was calculated following a bootstrap method. We performed $N$$\sim$1000 realisation where in each one we randomly drawn, without replacement, a parameter value pair $(A_w, \delta)$ from the output MCMC chain of our fitting method. In this way we also account for the degeneracies in the parameters of our model. The resulting normalised histograms of $r_0$ values from our bootstrap method are shown in the upper right corner inset plot of each panel in Figure~\ref{fig:Angular_Correlation_Function_zDependance_HATLAS_lowZ}. The black vertical dashed line indicates the mean of the distribution, which was derived by fitting a Gaussian distribution to the histogram. This value corresponds to our measurement of the clustering length, $r_0$ which is shown in the upper left corner of each panel, where the 1$\sigma$ uncertainty is taken as the standard deviation of the fitted Gaussian distribution. Our results are shown in the last column of Table~\ref{tab:Angular_Correlation_Function_zDependance_HATLAS_lowZ_Table} and seem to agree fairly well with \cite{vanKampen2012} measurements, even thought their uncertainties were considerable.

We need to note here that we find a significant difference in the measurement of the correlation length, $r_0$, in the redshift bin $0.15 < z < 0.2$ compared to \cite{vanKampen2012}. The authors report in their study the existence of a structure around $z\sim 0.164$, which might be responsible for the excess clustering strength. Due to the small area used in their analysis, this structure dominates the clustering signal in this redshift bin. However, we are using a much larger area in our study and this signal gets diluted, which is what probably causes this difference in the measurement of the clustering length.

\subsection{Clustering of $z>1$ SMGs} \label{sec:Clustering_highZ}

\begin{figure*}
\includegraphics[width=0.975\textwidth, height=0.375\textheight]{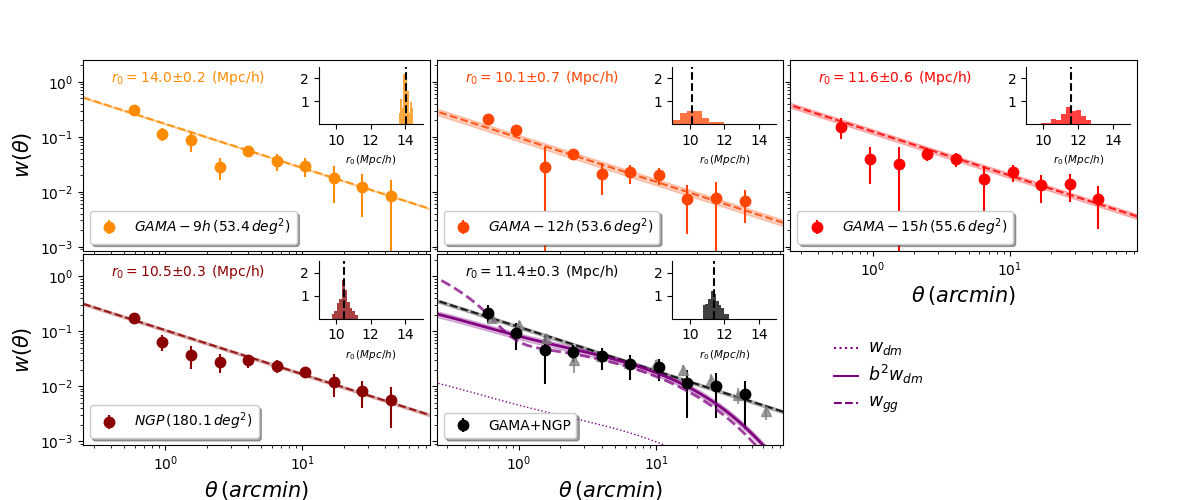} \\
\caption{The angular correlation function of sub-mm sources identified in the four \emph{H-ATLAS} fields: \textit{GAMA-9h} (top-left), \textit{GAMA-12h} (top-middle), \textit{GAMA-15h} (top-right) and \textit{NGP} (bottom-left). The error bars are derived using a 'delete one jackknife' resampling method \iffalse(described in Section~\ref{sec:Clustering_Analysis}) with the correlation matrix shown in Figure~\ref{fig:Covariance_Matrix}\fi. The bottom-middle panel shows the measured angular correlation function of the combined GAMA+NGP fields\iffalse, with the error bars given by Equation~\ref{eq:HATLAS_ACF_errors}\fi. \iffalse{\color{red}\fi The measurements were corrected by a factor of 1.25, as determined from our simulations in Appendix~\ref{sec:Appendix_A}, for the effect of filtering with a matched-filter.\iffalse}\fi The dashed line show the best-fit one parameter power-law model with fixed slope, $w(\theta)=A\theta^{-0.8}$, where the 1$\sigma$ uncertainty is shown as the shaded region. The inset plot in each panel shown the histogram of correlation length values which were derived from our bootstrap method. The black dashed vertical line in the inset plot of each panel, indicates the mean of the distribution which is also shown in the upper left corner in each panel. The purple dotted curve shows the dark matter angular correlation function, $w_{\textit{dm}}$. This has been scaled by the best-fit value of the linear bias factor, $b$, which is shown as the solid purple curve, with the 1$\sigma$ uncertainty shown as the shaded region. Finally, we show as the purple dashed curve the galaxy-galaxy angular correlation function, $w_{\textit{gg}}$, that corresponds to the best-fit HOD model. In addition, we show the results from {Gonz{\'a}lez-Nuevo} et al. (2017) as grey triangles.}
\label{fig:Angular_Correlation_Function_HATLAS_highZ}
\end{figure*}

Previous studies on the clustering of SMGs have focused on the broad redshift range $1<z<5$ \citep[e.g.][]{Webb2003, Blain2004, Scott2006, Weiss2009, Cooray2010, Maddox2010, Williams2011, Hickox2012, Mitchell-Wynne2012, Chen2016a, Chen2016b, Wilkinson2017}. As seen from the redshift distributions in Figure~\ref{fig:REDSHIFT_DISTRIBUTION_HATLAS_FIELDS}, the majority of our sources lie in that redshift range with the peak of the redshift distribution occurring around $z\sim1.25$ (excluding the optically identified counterparts which typically reside at $z<0.5$). Therefore, in order to make a direct comparison with previous clustering measurements, we first perform our clustering analysis for sources within this redshift range.

The measured angular correlation functions of sub-mm sources, for each of the \emph{H-ATLAS} fields under investigation, are shown in Figure~\ref{fig:Angular_Correlation_Function_HATLAS_highZ}: top panels for the three equatorial \emph{GAMA} fields and bottom-left for the \emph{NGP} field. Our measurements were corrected by a factor of 1.25, as determined from our simulations in Appendix~\ref{sec:Appendix_A}, for the effect of filtering with a matched-filter to remove the background cirrus emission. \iffalse}\fi The error bars were determined as $\sigma_i\sim\sqrt{C_{\textit{ii}}}$. 

In the bottom-middle panel of the same figure we show the measured angular correlation function by combining the three equatorial \emph{GAMA} fields with the \emph{NGP} field. In the same panel, we overlay the measurement from \cite{GonzalezNuevo2017} which was obtained using 250$\mu$m-selected sources in the redshift range $z>1.2$ from the \emph{GAMA} fields as well as a small part of the \emph{SGP} field. In this study the authors used sources selected with at least a 4$\sigma$ detection at 250$\mu m$, which results in a $S>29\,$mJy cut in flux density, and a 3$\sigma$ detection at 350$\mu m$ in order to preferentially select high redshift sources. The two measurements seem to agree fairly well across all angular scales. We can also compare our results with \cite{Cooray2010}, who used the two widest fields from the Herschel Multi-tiered Extragalactic Survey \citep[HerMES; ][]{Oliver2010}, Lockman-SWIRE and Spitzer First Look Survey (FLS). For this comparison there seems to be a large disagreement, with the authors of this paper reporting a stronger clustering signal across all angular scales. This disagreement, which was first realised by comparing the results from \cite{Maddox2010}, is alarming and it is not fully understood. We will discuss this further in Appendix~\ref{sec:Maddox_vs_Cooray}, where we suggest that the removal of the background cirrus emission being one possibility for this difference.

As a first step towards modelling the clustering signal, we use a one-parameter power-law model, $w(\theta) = A_w \theta^{-0.8}$, with a fixed slope. We perform the fitting for each individual field, as well as for the combined \emph{GAMA+NGP}. The resulting best-fit values for the parameter of our model are summarised in Table~\ref{tab:Angular_Correlation_Function_HATLAS_highZ_Table}. These correspond to the dashed colored lines in each panel of Figure~\ref{fig:Angular_Correlation_Function_HATLAS_highZ}, where the 1$\sigma$ uncertainty is shown as the shaded region.

The correlation length, $r_0$, was calculated following a bootstrap method in order to consider the uncertainty in the best-fit value of the power-law model. In each realisation we randomly sample the parameter $A_{\textit{w}}$ from a Gaussian distribution, centred at the best-fit value with a standard deviation equal to it's error, and use Equation~\ref{eq:Real_Space_CorrelationLength} to calculate the correlation length. The resulting normalised histograms of $r_0$ values, from our bootstrap method, are shown in the upper right corner of each panel in Figure~\ref{fig:Angular_Correlation_Function_HATLAS_highZ}. The black vertical dashed line indicates the mean of the distribution, which was derived by fitting a Gaussian distribution to the histogram. This value corresponds to our measurement of the clustering length, $r_0$ which is shown in the upper left corner of each panel, where the 1$\sigma$ uncertainty is taken as the standard deviation of the fitted Gaussian distribution. The results are summarized in Table~\ref{tab:Angular_Correlation_Function_HATLAS_highZ_Table}. We need to note here that the redshift distribution that enters the calculation of the correlation length, $r_0$, has been corrected for the effect of random photometric redshift errors, as described in Section~\ref{sec:SubmmPhotometricRedshiftsErrors}. 

We estimate the correlation length to be $r_0 = 11.4 \pm 0.4 \, h^{-1}$Mpc. The error in the measurement is relatively small, which is due to the assumption of a power-law model with a fixed slope thus reducing the uncertainties from introducing additional parameters. The measurement of the correlation length is in general agreement with previous studies \citep{Webb2003, Blain2004, Weiss2009, Williams2011, Hickox2012, Chen2016a, Chen2016b}. The measurement is also in agreement with \cite{Maddox2010}, who used 250$\mu$m-selected sources from the \emph{SDP} field of \emph{H-ATLAS}, reporting a clustering length in the range $r_0\sim7-11 \, h^{-1}$Mpc when considering additional colour cuts to preferentially select high-redshift sub-mm sources. However, comparing our results with \cite{Wilkinson2017} we seem to find a larger clustering strength, even when compared with their sample of SMGs with radio-identified counterparts which are typically comprised of more luminous SMGs.\iffalse}\fi

\subsubsection{Halo Bias model} \label{sec:halo_bias_model}

In order to convert the clustering strength to the inferred dark-matter halo mass, $M_{\textit{halo}}$, we need to compute the galaxy bias, $b$. This quantity can be inferred by scaling the dark-matter angular correlation function, $w_{dm}(\theta)$, according to the following relation:
\begin{equation} \label{eq:ACF_modelled}
w(\theta)=b^2 w_{dm}(\theta) \, .
\end{equation}
In the above expression the dark matter angular correlation function, $w_{dm}(\theta)$, can be computed using the Limber's equation which is used in order to convert a 3D power spectrum, $P(k)$ into a projected angular correlation function from,
\begin{equation}
w(\theta)= \frac{1}{c} \int \left( \frac{dN}{dz} \right)^2 H(z) \int \frac{k}{2\pi} P(k,z) J_0\left(\frac{k\theta}{\chi^{-1}(z)}\right) dk dz
\end{equation}
where $J_0$ is the zero-th order bessel function and $dN/dz$ is the corrected redshift distribution as described in Section~\ref{sec:SubmmPhotometricRedshiftsErrors}. In this case $P(k,z)$ is the non-linear dark matter power spectrum, $P_{\textit{NL}}(k,z)$ which was computed using the HALOMOD package (Murray et al. in prep). This package implements the HaloFit code (Smith et al. 2003) with improved parametrisation provided by Takahashi et al. (2012). 

Fitting our modelled angular correlation function, which is given by Equation~\ref{eq:ACF_modelled}, we determined the galaxy bias. Our theoretical prediction is shown as the purple curves in bottom-middle panel of Figure~\ref{fig:Angular_Correlation_Function_HATLAS_highZ}, where the 1$\sigma$ uncertainty is shown as the shaded region. The best-fit value of the galaxy bias for the combined \emph{GAMA+NGP} is $b=4.26\pm0.27$ (see Table~\ref{tab:Angular_Correlation_Function_HATLAS_highZ_Table}). 

Finally, in order to infer the dark matter halo mass that corresponds to a specific value of the galaxy bias we need to assume a bias function, $b(M,z)$. The value of the halo mass, $M_{\textit{halo}}$, will strongly depend on the assumed parametrisation of the bias function. We opted to use the function introduced by \cite{Tinker2010},
\begin{equation} \label{eq:bias}
b(\nu)=1-\frac{\nu^{\alpha}}{\nu^{\alpha} + \delta_c^{\alpha}} + B \nu^{b}+C \nu^{c}
\end{equation}
where $B=0.183$, $b=1.5$, $c=2.4$, $\delta_c$ is the critical density for collapse and $\nu=\delta_c/\sigma(M,z)$ is the "peak height" in the linear density field, with $\sigma(M,z)$ being the linear matter variance. 

The inferred dark matter halo mass using the bias function, which was detailed above, is $log(M_{\text{halo}})=13.2\pm 0.1$ (see Table~\ref{tab:Angular_Correlation_Function_HATLAS_highZ_Table}) and was calculated at the median redshift $\langle z \rangle \sim 1.75$. 

\subsubsection{Halo Occupation Distribution (HOD) model} \label{sec:halo_bias_model}

We can see for Figure~\ref{fig:Angular_Correlation_Function_HATLAS_highZ} that our model adopting the halo bias formalism does not provide an accurate fit to the small angular scales. In an attempt to model this clustering signal we make use of the halo model power spectrum, $P(k)$, which is written as the sum of two terms. The 1-halo term arises from interactions between galaxies within a single dark matter halo and dominates on small scales, while the 2-halo term arises from interactions of galaxies that belong to different halos and dominates on large scales \cite[see ][]{CooraySheth2002}. These terms are computed from,
\begin{equation}
P^{\textit{1h}}_{\textit{gg}}(k, z) = \int n(M,z) \frac{\langle N(N-1) | M \rangle}{\bar{N}^2_{\textit{gal}}} y^2(k|M,z) dM
\end{equation}
\begin{multline}  \label{eq:Power_Spectrum_mm_2h}
P^{\textit{2h}}_{\textit{gg}}(k, z) = P_{\textit{lin}}(k,z) \\
\left( \int n(M,z) b(M, z) \frac{\langle N | M \rangle}{\bar{N}_{\textit{gal}}} y(k|M,z) dM \right)^2 \, ,
\end{multline}
where $n(M,z)$ is the halo mass function \citep{Tinker2008}, $y(k|M,z)$ is the normalised Fourier transform of the halo density profile, $b(M,z)$ is the linear large-scale bias and $P_{\textit{lin}}(k,z)$ is the linear matter power spectrum which is computed using the \textit{CAMB} code \citep{LewisChallinorLasenby2002}.

This formalism introduces the Halo Occupation Distribution (HOD) parametrisation to the clustering signal arising from galaxy populations. In this parameterisation, the mean numbers of central and satellite galaxies in a halo of mass M are given by,
\begin{equation}
\langle N_{\textit{cen}} |M \rangle = \frac{1}{2} \left[ 1 + \textit{erf}\left( \frac{\textit{log}M - \textit{log}M_{\textit{cen}}}{\sigma_{\textit{log}M}} \right) \right] \, ,
\end{equation}
\begin{equation}
\langle N_{\textit{sat}} |M \rangle = \frac{1}{2} \left[ 1 + \textit{erf}\left( \frac{\textit{log}M - \textit{log}M_{\textit{cen}}}{\sigma_{\textit{log}M}} \right) \right] \left( \frac{M}{M_{\textit{sat}}} \right)^{\alpha_{\textit{sat}}} \, ,
\end{equation}
where $\textit{erf}(x)$ is the error function, $M_{\textit{cen}}$ is the minimum halo mass above which all halos host a central galaxy, $\sigma_{\textit{log}M}$ is the width of the central galaxy mean occupation, $M_{\textit{sat}}$ is the mass scale at which one satellite galaxy per halo is found, in addition to the central galaxy, and $\alpha_{\textit{sat}}$ is the power-law slope of the satellite occupation number with halo mass. 

The best-fit values of the parameters of our HOD model, which resulted from our MCMC analysis, are summarized in the first row of Table~\ref{tab:Angular_Correlation_Function_HATLAS_highZ_Table} for which we used flat priors for the parameters of our model within the range: $12<\log (M_{\textit{cen}}/h^{-1}M_{\odot})<14$, $10<\log (M_{\textit{sat}}/h^{-1}M_{\odot})<15$ with a fixed power-law slope for the satellite occupation number, $\alpha_{\textit{sat}}=1.0$, and width of the central galaxy mean occupation, $\sigma_{\textit{log}M}=0.3$.  Our theoretical prediction is shown as the purple dashed curve in bottom-middle panel of Figure~\ref{fig:Angular_Correlation_Function_zDependance_HATLAS} and seems to provide a more accurate fit to the data\iffalse}\fi. Using the Bayesian information criterion (BIC), we find: 
\begin{itemize}
\item $w(\theta)=A\theta^{-0.8}:$ 6.00 (1 parameter)
\item $w(\theta)=b^2 w_{\textit{dm}}\,:$ 6.32 (1 parameter)
\item $w(\theta)=w_{\textit{gg}}\,\,\,\,\,\,\,\,:$ 7.71 (2 parameter)
\end{itemize}
The BIC seem to prefer a power-law model, $w(\theta)=A\theta^{-0.8}$, which provides a better fit to the largest scales. If we were to exclude the last two data points, the BIC seem to prefer the $w(\theta)=w_{\textit{gg}}$ model.\iffalse}\fi

In addition, we would like to note that we considered using the HOD parametrisation of \cite{Geach2012}, which is more appropriate for star-formation rate selected samples. However, this HOD model has a lot more free parameters which were impossible to constrain given the errors in our measurements and the fact that we do not probe scales far in to the non-linear regime. The only parameter that is well constrained using this alternative parametrisation is the minimum halo mass above which all halos host a central galaxy $M_{\textit{cen}}$. This is the main parameter of interest for this work and it's value was consistent between the two different parameterisations.

\subsection{Evolution of Clustering with redshift for $z>1$ SMGs} \label{sec:Clustering_evolution_highZ}

\begin{figure}
\includegraphics[width=0.475\textwidth, height=0.325\textheight]{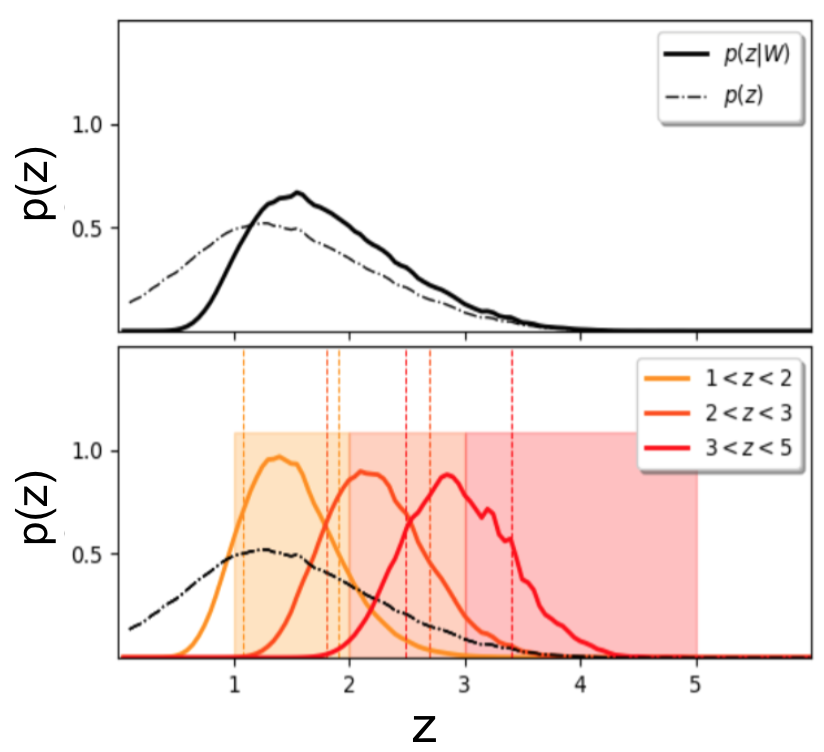} \\
\caption{The estimated redshift distributions $p(z|W)$ taking into account the window functions, $W(z_{\textit{ph}})$, and the photometric redshift error function, $p(z_{\textit{ph}}|z)$. The black dot-dashed line shows the initial redshift distribution, $p(z)$ of our sources. The top panel shows the "corrected" redshift distribution for sources in the redshift range $1<z<5$, while similarly in the bottom panel for the different redshift bins indicated at the right upper corner. The vertical solid lines correspond to the 50th percentile of the distribution, while the vertical dashed lines left and right of it correspond to the 16th and 84th percentiles respectively. The shaded regions show the width of our window functions.}
\label{fig:Corrected_Redshift_Distributions}
\end{figure}

The large sample of high-z sub-mm sources $(z>1)$ in the combined \emph{GAMA+NGP} fields allow us to investigate the redshift evolution of the clustering signal. To do that, we split our sample into three redshift bins, $1<z<2$, $2<z<3$ and $3<z<5$ similarly to \cite{Chen2016b}. The redshift distributions, $p(z|W)$, after accounting for the effect of random photometric redshifts are shown in last three panels of Figure~\ref{fig:Corrected_Redshift_Distributions} for the different redshift bins. We restricted our analysis to three redshift bins in order to avoid excessive overlap between the corrected redshift distribution.

\begin{figure*}
\includegraphics[width=0.9\textwidth, height=0.205\textheight]{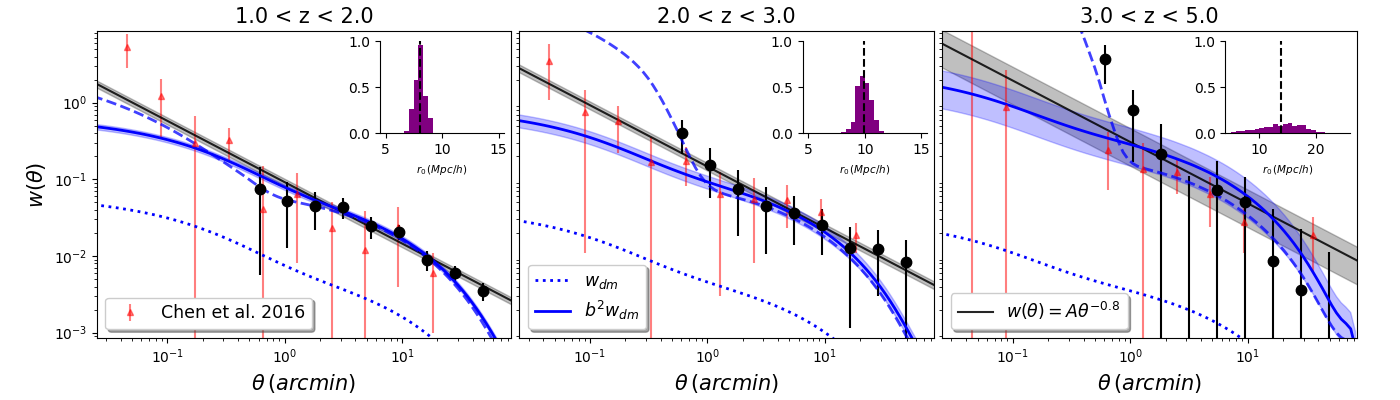} \\
\caption{The angular correlation function of sub-mm galaxies for each redshift slice in the redshift range $1<z<5$ (black circles).\iffalse, where the errors bars in this case are computed from Equation~\ref{eq:LandySzalayEstimator_Errors}.\fi \iffalse{\color{red}\fi The measurements were corrected by a factor of 1.25, as determined from our simulations in Appendix~\ref{sec:Appendix_A}.\iffalse}\fi The black solid lines corresponds to our fitted power-law model with a fixed slope, $w(\theta)=A\theta^{-0.8}$, where the 1$\sigma$ uncertainty is shown as the black shaded region. The inset plot in each panel show the histogram of correlation length values which were derived from our bootstrap method. The black dashed vertical line in the inset plot of each panel, indicates the mean of the distribution. The blue dotted curve shows the dark matter angular correlation function, $w_{\textit{dm}}$. This has been scaled by the best-fit value of the linear bias factor, b, which is shown as the solid blue curve, with the 1$\sigma$ uncertainty shown as the blue shaded region. Finally, we show as the blue dashed curve the galaxy-galaxy angular correlation function, $w_{\textit{gg}}$, that corresponds to the best-fit HOD model. In addition, we also include the measurements from Chen et al (2016) shown as red triangles.}
\label{fig:Angular_Correlation_Function_zDependance_HATLAS}
\end{figure*}

The resulting clustering measurements are shown in Figure~\ref{fig:Angular_Correlation_Function_zDependance_HATLAS} for each redshift bin. The measurements were corrected by a factor of 1.25 as determined from our simulations in Appendix~\ref{sec:Appendix_A}.\iffalse}\fi In each panel, we also include the measurement from \cite{Chen2016b} as red triangles, which probe angular scales down to $\sim$1". The two measurements agree fairly well in the angular scales probed by \emph{Herschel}. However, in the highest redshift bin we find an excess signal in the lowest probed angular bin compared to \cite{Chen2016b}. 

We fit a one-parameter power-law model with a fixed slope, $w(\theta) = A_w \theta^{-0.8}$, in order to model the angular correlation functions in each redshift bin. The resulting best-fit value for the parameter of our model, in each redshift bin, are shown in Table~\ref{tab:Angular_Correlation_Function_HATLAS_highZ_Table}. These corresponds to the black lines in each panel of Figure~\ref{fig:Angular_Correlation_Function_zDependance_HATLAS}, where the 1$\sigma$ uncertainty is shown as the grey-shaded region. 

\begin{table*}
	\centering{
	\caption{Results of Clustering Analysis for $z>1$ SMGs.}
	\begin{tabular}{lccccccccc} 
	\hline\\[-6.0mm]
  	\hline
	Sample & Field & $N_{\text{gal}}$ & $\langle z \rangle$ & $A_{\textit{w}}$ & $r_0$ & $b$ & $\textit{log}\left(\frac{M_{\textit{halo}}}{h^{-1}M_{\odot}}\right)$ & $\textit{log}\left(\frac{M_{\textit{cen}}}{h^{-1}M_{\odot}}\right)$ & $\textit{log}\left(\frac{M_{\textit{sat}}}{h^{-1}M_{\odot}}\right)$ \\
	& & & & & $(h^{-1}\textit{Mpc})$ & & & & \\
	\hline
	1$<$$z$$<$5 & \emph{GAMA+NGP} & 85319 & $1.75^{+0.55}_{-0.70}$ & $0.13 \pm 0.01$ & $11.4 \pm 0.4$ & $4.3 \pm 0.3$ & $13.2 \pm 0.1$ & $13.00^{+0.14}_{-0.22}$ & $14.17^{+0.65}_{-0.75}$ \\[0.25mm]
	\\
	1$<$$z$$<$2 & \emph{GAMA+NGP} & 55749 & $1.47^{+0.38}_{-0.46}$ & $0.09 \pm 0.01$ &  $8.1 \pm 0.5$ & $3.2 \pm 0.1$ & $13.1 \pm 0.1$ & $12.91^{+0.11}_{-0.14}$ & $14.78^{+0.50}_{-0.64}$ \\[0.25mm]
	\\
	2$<$$z$$<$3 & \emph{GAMA+NGP} & 25108 & $2.21^{+0.42}_{-0.49}$ & $0.11 \pm 0.02$ & $8.5 \pm 0.8$ & $4.5 \pm 0.4$ & $12.9 \pm 0.1$ & $12.50^{+0.28}_{-0.21}$ & $12.32^{+1.71}_{-1.29}$ \\[0.25mm]
	\\
	3$<$$z$$<$5 & \emph{GAMA+NGP} & 4462 & $2.93^{+0.45}_{-0.48}$ & $0.31 \pm 0.16$ & $13.9 \pm 3.9$ & $9.0 \pm 2.5$ & $13.2 \pm 0.2$ & $12.88^{+0.35}_{-0.41}$ & $12.93^{+1.71}_{-1.64}$ \\[0.25mm]
	\\
	\hline\\[-6.0mm]
	\hline
	\label{tab:Angular_Correlation_Function_HATLAS_highZ_Table}
	\end{tabular}
	\\
	}
\end{table*}

The correlation length, $r_0$, in each redshift slice was calculated following a bootstrap method, in order to consider the uncertainty in the best-fit value of the power-law model. In each realisation we randomly sample the parameter $A_{\textit{w}}$ from a Gaussian distribution, centred at the best-fit value with a standard deviation equal to it's error, and use Equation~\ref{eq:Real_Space_CorrelationLength} to calculate the correlation length. The resulting normalised histograms of $r_0$ values, from our bootstrap method, are shown in the upper right corner of each panel in Figure~\ref{fig:Angular_Correlation_Function_zDependance_HATLAS}. The black vertical dashed line indicates the mean of the distribution, which was derived by fitting a Gaussian distribution to the histogram. This value corresponds to our measurement of the clustering length, $r_0$, where the 1$\sigma$ uncertainty is taken as the standard deviation of the fitted Gaussian distribution. Our results are shown in Table~\ref{tab:Angular_Correlation_Function_HATLAS_highZ_Table} for each redshift slice.
\iffalse}\fi

Finally, we compute the bias parameters, $b$, for each redshift slice following the same methodology outlined in Section~\ref{sec:Clustering_highZ}, using the corrected redshift distributions shown in Figure~\ref{fig:Corrected_Redshift_Distributions} in order to compute the projected the dark matter angular correlation functions, $w_{\textit{dm}}(\theta)$. In Figure~\ref{fig:Angular_Correlation_Function_zDependance_HATLAS} we shown $w_{\textit{dm}}(\theta)$ in each panel as the blue dashed lines. In the same Figure, the blue solid lines in each panel show the projected the dark matter angular correlation functions scaled by the best-fit value of the linear bias parameters, where the 1$\sigma$ uncertainty is shown as the blue-shaded region. Our results are shown in Table~\ref{tab:Angular_Correlation_Function_HATLAS_highZ_Table} for each redshift slice along with the halo masses, $M_{\textit{halo}}$, that correspond to these bias measurements according to Equation~\ref{eq:bias}.

In the last two panels of Figure~\ref{fig:Angular_Correlation_Function_zDependance_HATLAS} we see that the scaled dark matter angular correlation functions does not provide a satisfactory fit to the data, indicating the need of using an HOD model, similar to the analysis in the previous section. The results from our MCMC analysis are shown in Table~\ref{tab:Angular_Correlation_Function_HATLAS_highZ_Table} for which we used flat priors for the parameters of our model within the range: $12<\log (M_{\textit{cen}}/h^{-1}M_{\odot})<14$ and $10<\log (M_{\textit{sat}}/h^{-1}M_{\odot})<15$ with a fixed power-law slope for the satellite occupation number, $\alpha_{\textit{sat}}=1.0$, and width of the central galaxy mean occupation, $\sigma_{\textit{log}M}=0.3$. We were not able to set good constrains on $M_{\textit{sat}}$. The resulting errors depend strongly on the range of prior, we adopted for this parameter.


\section{Discussion}

\begin{figure*}
\includegraphics[width=0.95\textwidth, height=0.45\textheight]{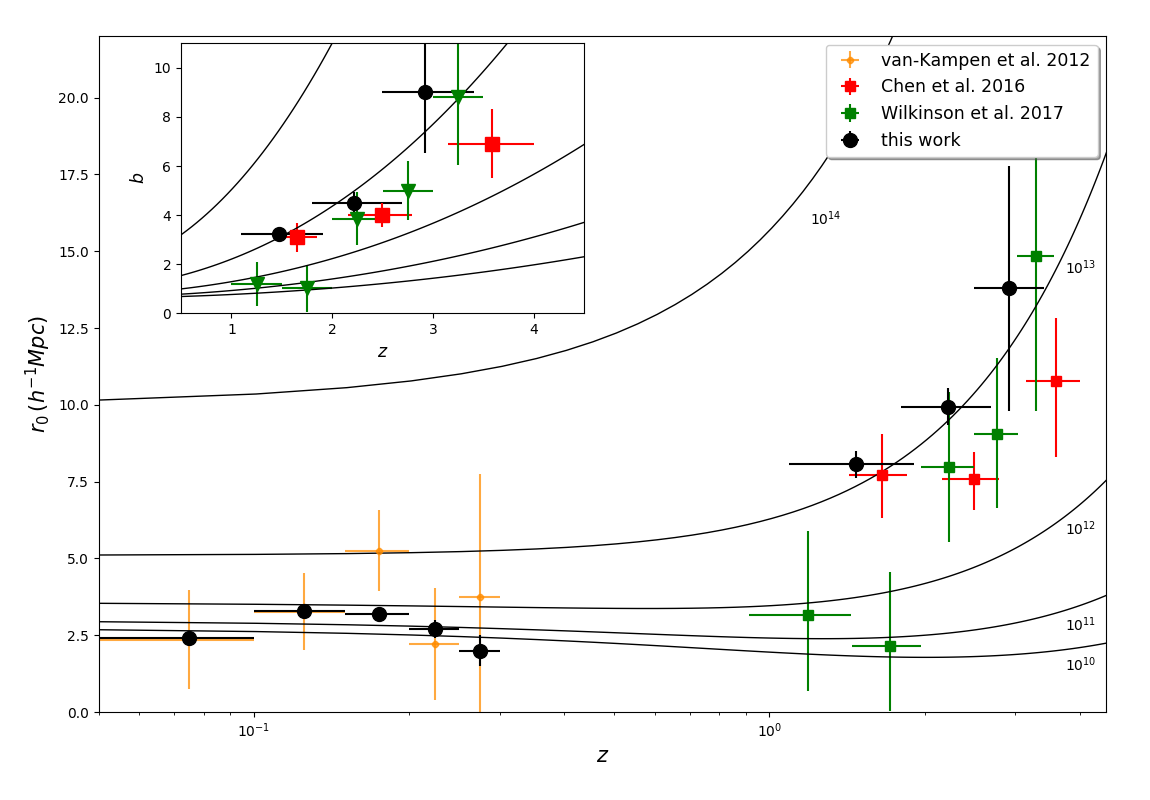} \\
\caption{The evolution of the correlation length $r_0$ with redshift for our sample of 250$\mu$m-selected sources with flux densities $S>30\,$mJy (black points). We also show the clustering results from previous studies: Herschel-ATLAS science demonstration phase (SDP) field 250$\mu$m-selected sources at 0.05 < z < 0.3 (van-Kampen et al. 2012; yellow points), UKIRT Infrared Deep Sky Survey (UKIDSS) 850$\mu$m-selected SMGs at 1 < z < 5 (Chen et al. 2016b; red points), SCUBA-2 Cosmology Legacy Survey 850$\mu$m-selected SMG's at 1 < z < 3.5 (Wilkinson et al. 2017; green points). The black solid lines show the evolution of $r_0$ with redshift for dark matter halos of different masses (in units of $h^{-1}M_{\odot}$) using Equation~\ref{eq:corr_length}. The inset plot of the top left corner shows the evolution of the galaxy bias as a function of redshift, where the black solid lines show the theoretical predictions using Equation~\ref{eq:bias} of the bias function from Tinker et al. (2010).}
\label{fig:Correlation_Length_Evolution}
\end{figure*}

Our findings are summarised in Figure~\ref{fig:Correlation_Length_Evolution} where we plot the evolution of the correlation length, $r_0$, as a function of redshift for our sample of 250$\mu$m-selected sources with flux densities $S>30\,$mJy. The green points correspond to measurements from \cite{Wilkinson2017}, while the red points correspond to measurements from \cite{Chen2016b}. The black lines are the theoretical predictions for the evolution of the correlation length with redshift for different halo masses, which were estimated using the formalism of \cite{Peebles1980}. According to that formalism the correlation length is related to the bias parameter, as 
\begin{equation} \label{eq:corr_length}
r_0=8 \left( \frac{\Delta_8^2}{C_{\gamma}} \right)^{1/\gamma} = 8 \left( \frac{b^2\sigma_8^2 D^2}{C_{\gamma}} \right)^{1/\gamma}
\end{equation} 
where $\Delta_8$ is the clustering strength of haloes, more massive than the mass $M$ at redshift $z$ and is defined as  $\Delta_8(M,z)=b(M,z)\sigma_8D(z)$, with $D(z)$ being the growth factor of linear fluctuations in the dark matter distribution which is computed from,
\begin{equation}
D(z) = \frac{5 \Omega_m E(z)}{2} \int_z^{\infty} \frac{1 + y}{E^3(y)} dy \, .
\end{equation}
The factor $C_{\gamma}$ is computed from, $C_{\gamma} = 72/(3 - \gamma)(4 - \gamma)(6 - \gamma)2^{\gamma}$, 
where $\gamma$ is the slope of power-law model which parametrises the spatial correlation function and is taken to be $\gamma=1.8$ (since we assume the same power-law slope when computing the correlation length, see Section~\ref{sec:correlation_length}). The inset plot in Figure~\ref{fig:Correlation_Length_Evolution} shows the evolution of the bias parameter as a function of redshift, where the green and red points correspond to the values found in the aforementioned studies. The black solid lines correspond to theoretical predictions using Equation~\ref{eq:bias} of the bias function from \cite{Tinker2010}.

At high redshifts our results are in general agreement with previous studies for the evolution of clustering of SMGs. Above redshift of about $\sim2$, however, we find our population of bright SMGs selected at 250$\mu$m with flux densities $S_{250}>30$mJy exhibit larger clustering strengths (2$\,\sigma$ discrepancy) compared to \cite{Chen2016b} where the authors studied a sample of faint SMGs selected at 850$\mu$m with flux densities $S_{850}<2$mJy. This indicates that brighter SMGs cluster more strongly than their faint counterparts even at high redshifts, which is also supported by the fact that our results are more in agreement with \cite{Wilkinson2017} where the authors studied bright SMGs selected at 850$\mu$m with flux densities $S_{850}>2$mJy. On the other hand, we find that SMGs in the redshift range $1<z<2$ follow the same evolutionary track as those at higher redshifts, in contrast to the findings of \cite{Wilkinson2017} where the authors reported a downsizing effect (3$\,\sigma$ discrepancy). However, this effect is not present in the analysis of \cite{Chen2016b}. It is not straightforward to determine the cause for this difference, as there might be biases folded in the measurements associated with the selection of these SMGs.

The discrepancies of the aforementioned differences are at 2-3$\,\sigma$. If we also were to consider that the errors on our measurements are slightly underestimated as determined from our simulations in Appendix~\ref{sec:Appendix_A}, the agreement becomes even better. This suggest that we need to improve the accuracy of our measurements in order to confidently differentiate the clustering properties of faint and bright SMGs, as well as SMGs selected at different wavelengths. This improvement can come by obtaining more realistic photometric redshift measurements and potentially use the Atacama Large Millimeter/submillimeter Array (ALMA) to improve our counterpart identification techniques \citep{Jin2018}.\iffalse}\fi
\iffalse}\fi
 

\section{Conclusions}

We measured the angular auto-correlation function of low- and high-redshift sub-mm sources identified in the \emph{GAMA+NGP} fields of the \emph{H-ATLAS}, which comprise the largest area extragalactic survey at sub-mm wavelengths. We selected a sample of sources detected at the 4$\sigma$ significance level ($\sim$ 30 mJy) at 250-$\mu$m. Our main results are summarised as follows:

\begin{itemize}
\item We performed simulations of clustered sources and assessed our methodology of extracting sources from our 250$\mu$m \emph{H-ATLAS} maps. We estimated the correction factor that needs to be applied to the measured angular correlation function due to the loss of clustering power from our method for removing Galactic cirrus emission. Our simulations and methodology for calculating this correction factor are described in Appendix A.  
\item First, we studied the evolution of clustering with redshift for our low redshift $(z<0.3)$ sample. We showed that SMGs in this redshift range exhibit clustering lengths of the order of $\sim 2-3 h^{-1}$Mpc, similar to normal galaxies selected at optical wavelengths. Our results agree with the findings of \cite{vanKampen2012}, albeit with much improved errors on the measurement due to our larger sample.
\item We performed an auto-correlation analysis of SMG in the redshift range $1<z<5$, which is similar to the redshift range of many previous studies. We showed that SMGs are strongly clustered, finding a clustering length $r_0=11.4 \pm 0.4 h^{-1}$Mpc. We modelled the clustering signal by scaling the dark matter angular correlation function by the linear bias factor, finding a value of $b=4.26\pm0.27$ that corresponds to a dark matter halo mass of $\log(M_{\textit{halo}})=13.2\pm0.1h^{-1}M_{\odot}$.
\item In addition, we studied the evolution of clustering with redshift for our sample of high redshift $(z>1)$ sub-mm sources. We showed that SMGs occupy dark matter halos with masses of the order of $10^{12.9}$, $10^{12.5}$ and $10^{12.9}h^{-1}M_{\odot}$ at $z=1-2$, $2-3$ and $3-5$, respectively. We did not find a downsizing effect for SMG below redshift of about $\sim2$, as reported in \cite{Wilkinson2017}.

\item Finally, we point out that galaxies selected at 250$\mu$m at high and low redshifts are not the same population. The star formation activity seems to be shifting from high mass halos at $z>1$ to less massive halos at $z<1$, consistent with the downsizing effect reported in \cite{Magliocchetti2014}.
\end{itemize}


\section*{Acknowledgements}

We thank the anonymous referee for the insightful comments and suggestions which helped to improve this manuscript.
EV and SAE acknowledge funding from the UK Science and Technology Fa- cilities Council consolidated grant ST/K000926/1. JGN acknowledges financial support from the I+D 2015 project AYA2015- 65887-P (MINECO/FEDER) and from the Spanish MINECO for a "Ramon y Cajal" fellowship (RYC-2013-13256). MN acknowledges financial support from the European Union's Horizon 2020 research and innovation programme under the Marie Sk{\l}odowska-Curie grant agreement No 707601. MS and SAE have received funding from the European Union Seventh Frame- work Programme ([FP7/2007-2013] [FP7/2007-2011]) under grant agreement No. 607254. LD and SM acknowledge support from the European Research Council (ERC) in the form of Consolidator Grant COSMICDUST (ERC-2014-CoG-647939, PI H L Gomez). GDZ acknowledges support from ASI/INAF agreement n.~2014-024-R.1  and from the ASI/Physics Department of the university of Roma--Tor Vergata agreement n. 2016-24-H.0.



\appendix


\section{Correction to the Angular Correlation Function} \label{sec:Appendix_A}

\begin{figure*}
\includegraphics[width=0.65\textwidth, height=0.40\textheight]{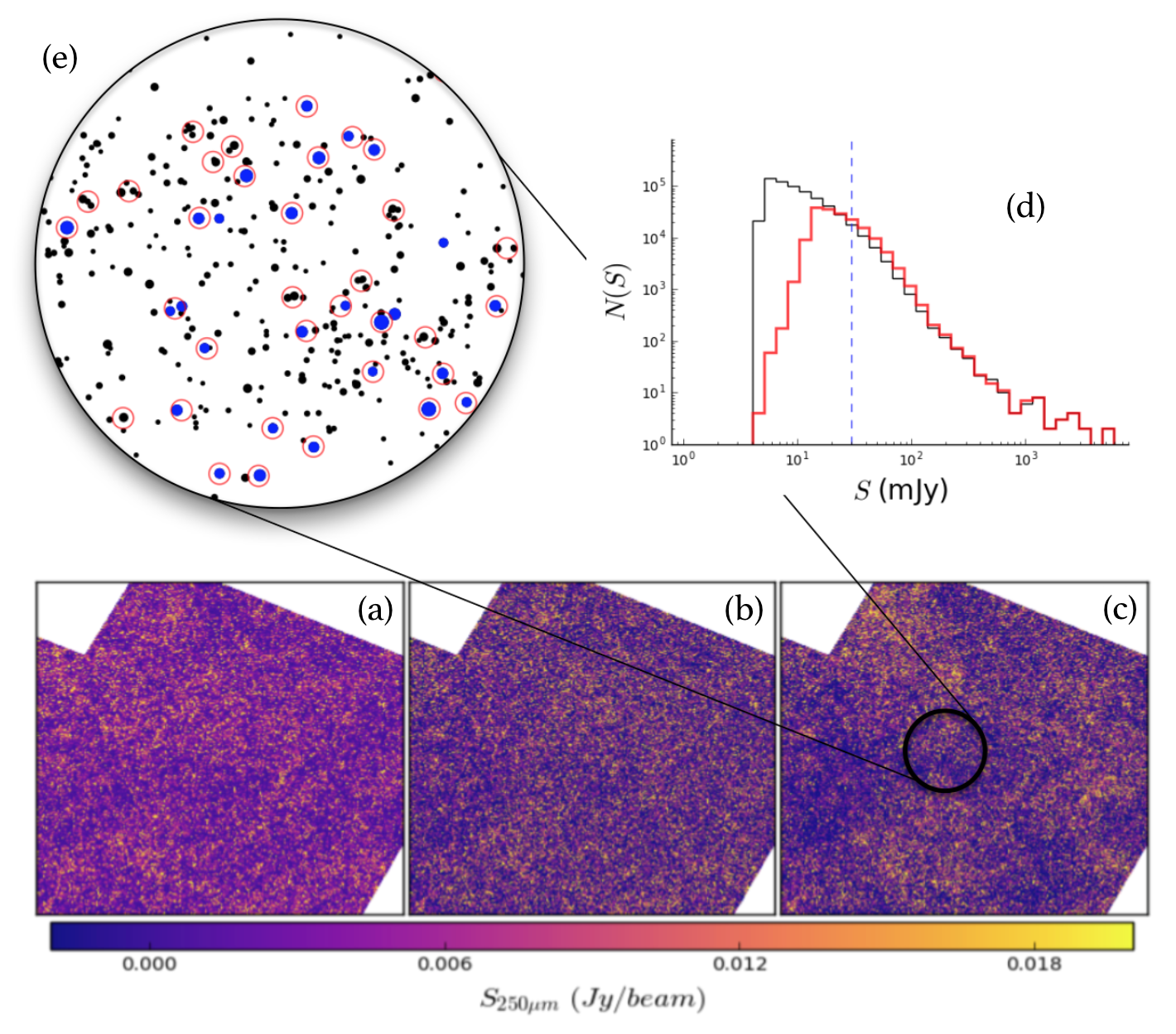} \\
\caption{The different steps illustrating the procedure followed to produce realistic maps of clustered point sources for the \textit{GAMA}-$15h$ field (see text for more details). Panel (a) shows the convolved map of our input point sources, panel (b) the noise was added and in panel (c) the background emission from the \emph{Nebuliser} using a pixel scale $N_{\textit{pix,b}} = 30\,$arcmin was added and the mean of the map was set to zero. Applying our source extraction algorithm (MADX) using a pixel scale $N_{\textit{pix,f}} = 30\,$arcmin we obtain the number counts as a function of flux density shown as the red histogram in panel (d), with the input number counts shown as the black histogram respectively. Finally, panel (e) shows a zoomed in region of panel (c) where the black and blue points correspond to the input sources, with the blue being the sources with flux densities >30mJy. The red circles on the other hand correspond to our extracted sources with flux densities >30mJy.}
\label{fig:SIMULATED_MAP_EXAMPLE}
\end{figure*}

One concern that was pointed out by \cite{Maddox2010} is the effect of removing the Galactic cirrus emission, on the measured clustering signal of submillimeter sources. This emission was removed, using the \emph{Nebuliser} algorithm, before trying to detect extragalactic sources with the MADX (Maddox et al. in preparation) algorithm on the \emph{Herschel} images (as discussed in Section~\ref{sec:H-ATLAS_DATA}). In addition to the background cirrus emission, \emph{Nebuliser} can also remove any large scale background produced by clustered faint sources that cannot be individually resolved and can ultimately affect the measured clustering signal \citep{Valiante2016,Maddox2018}. 

\subsection{Generating Realistic Maps}

In order to quantify this effect we create catalogues of clustered source positions on the sky region covered by the \textit{GAMA}-$15h$ field according to an input power-law power spectrum $P_{\textit{corr}}(k)$\footnote{In the case of a power-law power spectrum $P(k) \propto k^{-1.2}$, the angular correlation function $w(\theta) \propto \theta^{-0.8}$}. For these catalogues we follow the methodology of \cite{GonzalezNuevo2005} (G05). We adopt the model by \cite{Cai2013} for the number count distribution of these sources assuming a minimum flux limit of $S_{\textit{min}} = 5 \,$mJy, unless otherwise stated (we refer the reader to the G05 paper for more details). 

These catalogues were then used to produce realistic maps of the \textit{GAMA}-$15h$ field. We start by creating a high resolution map (1$"$/pixel) on top of which our simulated sources are laid down. This map is convolved with a PSF (the measured FWHM of the azimuthially average circular PSF is 17.8 at 250$\mu$m) and consequently rescaled to the real \emph{Herschel} pixel size at 250 microns (6$"$/pixel). In panel (a) of Figure~\ref{fig:SIMULATED_MAP_EXAMPLE} a small patch of this map is shown. Subsequently, the noise map was added to the image. This map was created by assigning each pixel a value drawn from a gaussian distribution with zero mean and standard deviation equal to the corresponding pixel value of the raw instrumental noise map of the \textit{GAMA}-$15h$ field \citep[see ][]{Valiante2016}. In panel (b), the same patch of the map is shown after the noise was added. Finally, we included the Galactic background cirrus emission that was estimated by \emph{Nebulizer} on our real Herschel image of the \textit{GAMA}-$15h$ field, using a filtering scale of $N_{\textit{pix}}$ pixels. We will use the notation $N_{\textit{pix,b}}$ when we refer to the background that was added to our simulated map. The last step of the map-making procedure is to set the mean of the map to zero. In panel (c) of Figure~\ref{fig:SIMULATED_MAP_EXAMPLE} we show the case with $N_{\textit{pix,b}} = 30$.

\begin{figure*}
\includegraphics[width=0.95\textwidth, height=0.175\textheight]{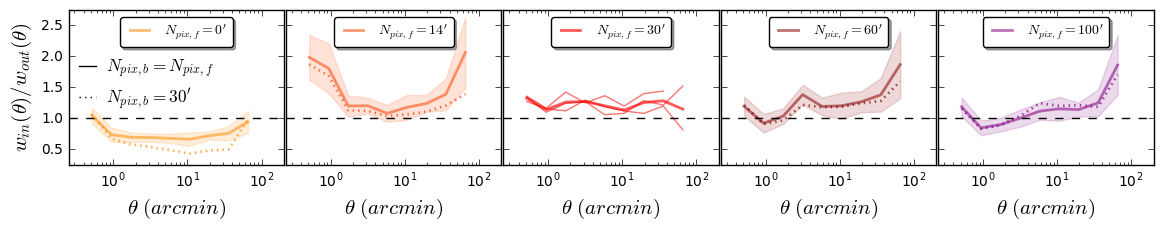} \\
\caption{The ratio of the measured input to the output (after filtering) angular correlation function. The different panels, going from left to right, correspond to a filtering scale of $N_{\textit{pix,f}} = 0, 14, 30, 60, 100$. In all panels, the dotted lines correspond to a added background emission that was produced by the \emph{Nebuliser} algorithm using a filtering scale $N_{\textit{pix,b}} = 30$. The continuous lines, on the other hand, correspond to a added background emission using $N_{\textit{pix,b}} = N_{\textit{pix,f}}$, where the shaded regions are the 1$\sigma$ Poisson uncertainties to the measurements. In the middle panel, corresponding to a filtering scale $N_{\textit{pix,b}} = 30$, we performed this procedure for three realisations of our simulations. These are shown as the thin faded continuous lines where the thick line shows their average.}
\label{fig:ANGULAR_CORRELATION_FUNCTION_SIMULATION}
\end{figure*}

Once our simulated map is created we then execute our source extraction algorithm after we filter our map with Nebulizer with a scale in pixels equal to $N_{\textit{pix,f}}$. We will use the notation $N_{\textit{pix,f}}$ when we refer to the filtering of our simulated maps prior to source extraction. In Figure~\ref{fig:SIMULATED_MAP_EXAMPLE} we have used $N_{\textit{pix,b}} = N_{\textit{pix,f}} = 30$. In the right top corner we show the number counts of our input catalogue of simulated point sources as the black histogram along with the number counts of sources extracted from our simulated map as the red histogram. For a visual aid in the left top corner we show the input and output source catalogues in a smaller region of the map. The black and blue points correspond to our input sources (the size of these points is indicative of their flux densities), where blue points are sources with flux densities $S\ge 30\,$mJy. Equivalently, the open red circles are sources in the output catalogue with flux densities $S\ge 30\,$mJy. As shown from this Figure, as well as from the histograms, there is a clear flux boosting effect taking place due to the confusion of low flux density sources.

\subsection{Determining the correction factor}

In order to now quantify the effect in question we compute the ratio of input to the output angular correlation function for sources with $S > 30\,$mJy, where the results are shown in Figure~\ref{fig:ANGULAR_CORRELATION_FUNCTION_SIMULATION}. The different panels correspond to the different filtering pixel scales used in the analysis $N_{\textit{pix,f}} = 0, 14, 30, 60, 100$ starting from left to right, respectively. The continuous lines correspond to a background with $N_{\textit{pix,b}} = N_{\textit{pix,f}}$ pixel scale, while for the dotted lines we used the same background for all panel of $N_{\textit{pix,b}} = 30$. The case of $N_{\textit{pix,b}} = 0$ means that no background emission was added.

In the case where no filtering is being applied prior to the source extraction (first panel of Figure~\ref{fig:ANGULAR_CORRELATION_FUNCTION_SIMULATION}) the resulting clustering strength of our extracted sources is enhanced in all angular scales, whether or not a background is added. This results from the combination of two effects: (i) The low flux density sources of our input catalogue, which constitute the unresolved background, are also clustered. (ii) The added background cirrus emission is also contributing to the enchantment of the clustering signal. In the remaining cases, we filter our simulated map prior to the extraction of sources using the different filtering pixel scales. The aim of filtering the map is to remove the cirrus emission only, but as a consequence some real clustering is also removed.

We determine that the case of $N_{\textit{pix,f}} = 30$, which is in fact the one used for extracting sources from the H-ATLAS maps, performs best. From the ratio of input to output angular correlation functions we work out that a 25\% correction needs to be applied, at all angular scales, to the clustering measurements. In fact, for this case only we performed this procedure for three realisations of our simulation, which are shown as the thin faded red lines in the middle panel of Figure~\ref{fig:ANGULAR_CORRELATION_FUNCTION_SIMULATION}, where the thicker red line is their average. We use this correction factor in our analysis, only when specifically stated. \iffalse}\fi  In addition, we see from the different realisations that there is non-negligible scatter, which will contribute to the final error budget of the measured clustering properties presented in Figure~\ref{fig:Correlation_Length_Evolution}. As a result the quoted errors in Figure~\ref{fig:Correlation_Length_Evolution} are slightly underestimated.\iffalse}\fi

\subsection{Maddox et al. (2010) vs Cooray et al. (2010)}\label{sec:Maddox_vs_Cooray}

We pointed out in Section~\ref{sec:Clustering_highZ} that our measurement of the angular correlation function is significantly lower compared to the measurement of \cite{Cooray2010}. In their case, no filtering was applied prior to source extraction, even though there should be cirrus contamination to some degree in the HERMES fields. This means that their measurement would fall into the case of $N_{\textit{pix,b}}=30$ and $N_{\textit{pix,f}}=0$. On the other hand, our measurement would corresponds to $N_{\textit{pix,b}}=30$ and $N_{\textit{pix,f}}=30$. Our simulations in Figure~\ref{fig:ANGULAR_CORRELATION_FUNCTION_SIMULATION} suggest that there is more than a 50\% difference between these two cases. This could be the reason of the disagreement between the two measurements, which was first pointed out by \cite{Maddox2010}.

\bsp	
\label{lastpage}

\begin{thebibliography}{99}
\bibitem[\protect\citeauthoryear{Abazajian et al.}{2009}]{Abazajian2009}
Abazajian N. K., Adelman-McCarthy J. K., Agueros M. A., et al., \href{http://adsabs.harvard.edu/abs/2009ApJS..182..543A}{2009},  ApJS, 182, 543 
\bibitem[\protect\citeauthoryear{Adelberger et al.}{2005}]{Adelberger2005}
{Adelberger}, K.~L., {Steidel}, C.~C., {Pettini}, M., {Shapley}, A.~E., et al., \href{http://adsabs.harvard.edu/abs/2005ApJ...619..697A}{2005}, ApJ, 619, 697 
\bibitem[\protect\citeauthoryear{Almeida, Baugh \& Lacey}{2011}]{Almeida2011}
{Almeida}, C., {Baugh}, C.~M., {Lacey}, C.~G., \href{http://adsabs.harvard.edu/abs/2011MNRAS.417.2057A}{2011}, MNRAS, 417, 2057
\bibitem[\protect\citeauthoryear{Amblard et al.}{2011}]{Amblard2011}
{Amblard}, A., {Cooray}, A., {Serra}, P., {Altieri}, B., {Arumugam}, et al., \href{http://adsabs.harvard.edu/abs/2011Natur.470..510A}, Nature, 470, 510
\bibitem[\protect\citeauthoryear{Bakx et al.}{2018}]{Bakx2018}
{Bakx}, T.~J.~L.~C., {Eales}, S.~A., {Negrello}, M., {Smith}, M.~W.~L., {Valiante}, et al., \href{http://adsabs.harvard.edu/abs/2018MNRAS.473.1751B}{2018}, MNRAS, 473, 1751
\bibitem[\protect\citeauthoryear{Blain et al.}{2002}]{Blain2002}
Blain A. W., Smail I., Ivison R. J., Kneib J.-P., Frayer D. T, \href{http://adsabs.harvard.edu/abs/2002PhR...369..111B}{2002}, Phys. Rep., 369, 111
\bibitem[\protect\citeauthoryear{Blain et al.}{2004}]{Blain2004}
Blain A. W., Chapman S. C., Smail I., Ivison R., \href{http://adsabs.harvard.edu/abs/2004ApJ...611..725B}{2004}, ApJ, 611, 725
\bibitem[\protect\citeauthoryear{Bourne et al.}{2014}]{Bourne2014}
{Bourne}, N., {Maddox}, S.~J., {Dunne}, L., {Dye}, S., {Eales}, S, et al., \href{http://adsabs.harvard.edu/abs/2014MNRAS.444.1884B}{2014}, MNRAS, 444, 1884
\bibitem[\protect\citeauthoryear{Bourne et al.}{2016}]{Bourne2016}
Bourne N., Dunne L., Maddox S. J., Dye S., Furlanetto C., et al., \href{http://adsabs.harvard.edu/abs/2016MNRAS.462.1714B}{2016}, MNRAS, 462, 1714
\bibitem[\protect\citeauthoryear{Brodwin et al.}{2008}]{Brodwin2008}
Brodwin M., Dey A., Brown M. J. I., Pope A., et al., \href{http://adsabs.harvard.edu/abs/2008ApJ...687L..65B}{2008}, ApJ, 687, 65
\bibitem[\protect\citeauthoryear{Bryan \& Norman}{1998}]{BryanNorman1998}
Bryan, G. L., Norman M. L., \href{http://adsabs.harvard.edu/abs/1998ApJ...495...80B}{1998}, ApJ, 495, 80
\bibitem[\protect\citeauthoryear{{Budav{\'a}ri} et al.}{2003}]{Budavari2003}
{Budav{\'a}ri}, T., {Connolly}, A.~J., {Szalay}, A.~S., {Szapudi}, I., {Csabai}, I., et al., \href{http://adsabs.harvard.edu/abs/2003ApJ...595...59B}{2003}, ApJ, 595, 59
\bibitem[\protect\citeauthoryear{Bussmann et al.}{2013}]{Bussmann2013}
{Bussmann}, R.~S., {P{\'e}rez-Fournon}, I., {Amber}, S., {Calanog}, J., {Gurwell}, M.~A., et al., \href{http://adsabs.harvard.edu/abs/2013ApJ...779...25B}{2013}, ApJ, 779, 25
\bibitem[\protect\citeauthoryear{Bullock et al.}{2001}]{Bullock2001}
Bullock J. S., Kolatt T. S., Sigad Y., Somerville R. S., Kravtsov A. V., et al., \href{http://adsabs.harvard.edu/abs/2001MNRAS.321..559B}{2001}, MNRAS, 321, 559
\bibitem[\protect\citeauthoryear{Cai et al.}{2013}]{Cai2013}
Cai Z.-Y., Lapi A., Xia J.-Q., De Zotti G., Negrello M., Gruppioni C., et al., \href{http://adsabs.harvard.edu/abs/2013ApJ...768...21C}{2013}, ApJ, 768, 21
\bibitem[\protect\citeauthoryear{Casey et al.}{2014}]{Casey2014}
Casey C. M., Narayanan D., Cooray A., \href{http://adsabs.harvard.edu/abs/2014PhR...541...45C}{2002}, Phys. Rev., 541, 45
\bibitem[\protect\citeauthoryear{Chapin et al.}{2011}]{Chapin2011}
Chapin E. L., Chapman S. C., Coppin K. E., Devlin M. J., et al., \href{http://adsabs.harvard.edu/abs/2011MNRAS.411..505C}{2011}, MNRAS, 411, 505
\bibitem[\protect\citeauthoryear{Chapman et al.}{2005}]{Chapman2005}
Chapman S. C., Blain A. W., Smail I., Ivison R. J., \href{http://adsabs.harvard.edu/abs/2005ApJ...622..772C}{2005}, ApJ, 622, 772
\bibitem[\protect\citeauthoryear{Chen et al.}{2016a}]{Chen2016a}
Chen C.-C., Smail I., IvisonR. J., Arumugam V., Almaini O., et al., \href{http://adsabs.harvard.edu/abs/2016ApJ...820...82C}{2016a}, ApJ, 820, 82
\bibitem[\protect\citeauthoryear{Chen et al.}{2016b}]{Chen2016b}
Chen C.-C., Smail I., Swinbank A. M., Simpson J. M., et al., \href{http://adsabs.harvard.edu/abs/2016ApJ...831...91C}{2016b}, ApJ, 831, 91
\bibitem[\protect\citeauthoryear{Colless et al.}{2001}]{Colless2001}
Colless M., Dalton G., Maddox S., Sutherland W., Norberg P. Cole S., et al., \href{http://adsabs.harvard.edu/abs/2001MNRAS.328.1039C}{2001}, MNRAS, 328, 1039
\bibitem[\protect\citeauthoryear{Cooray \& Sheth}{2002}]{CooraySheth2002}
Cooray A., Sheth R., \href{http://adsabs.harvard.edu/abs/2002PhR...372....1C}{2002}, Phys. Rev., 372, 1
\bibitem[\protect\citeauthoryear{Cooray et al.}{2010}]{Cooray2010}
Cooray A., Amblard A., Wang L., Arumugam V., Auld R., et al., \href{http://adsabs.harvard.edu/abs/2010A\%26A...518L..22C}{2010}, A\&A, 518, 22
\bibitem[\protect\citeauthoryear{Coppin et al.}{2006}]{Coppin2006}
Coppin K., Chapin E. L., Mortier A. M. J., Scott S. E., et al., \href{http://adsabs.harvard.edu/abs/2006MNRAS.372.1621C}{2006}, MNRAS, 372, 1621
\bibitem[\protect\citeauthoryear{Cox et al.}{2011}]{Cox2011}
{Cox}, P., {Krips}, M., {Neri}, R., {Omont}, A., {G{\"u}sten}, R., {Menten}, K.~M., et al., \href{http://adsabs.harvard.edu/abs/2011ApJ...740...63C}{2011}, ApJ, 740, 63
\bibitem[\protect\citeauthoryear{Cowley et al.}{2017}]{Cowley2017}
{Cowley}, W.~I., {Lacey}, C.~G., {Baugh}, C.~M., {Cole}, S., {Wilkinson}, A., \href{http://adsabs.harvard.edu/abs/2017MNRAS.469.3396C}{2017}, MNRAS, 469, 3396
\bibitem[\protect\citeauthoryear{{Dav{\'e}} et al.}{2010}]{Dave2010}
{Dav{\'e}}, R., {Finlator}, K., {Oppenheimer}, B.~D., {Fardal}, et al., \href{http://adsabs.harvard.edu/abs/2010MNRAS.404.1355D}{2010}, MNRAS, 404, 1355
\bibitem[\protect\citeauthoryear{de Jong et al.}{2015}]{de_Jong2015}
de Jong J. T. A., Verdoes K., G. A., Boxhoorn D. R., Buddelmeijer H., Capaccioli M., et al., \href{http://adsabs.harvard.edu/abs/2015A\%26A...582A..62D}{2015}, A\&A, 582, A62
\bibitem[\protect\citeauthoryear{Driver et al.}{2009}]{Driver2009}
{Driver}, S.~P., {GAMA Team}, {Baldry}, I.~K., {Bamford}, S., {Bland-Hawthorn}, J., {Bridges}, T., et al., \href{http://adsabs.harvard.edu/abs/2009IAUS..254..469D}{2009}, Astron. Geophys., 50, 12
\bibitem[\protect\citeauthoryear{Eales et al.}{2010}]{Eales2010}
Eales S., Dunne L., Clements D., Cooray A., De Zotti G., Dye S., et al., \href{http://adsabs.harvard.edu/abs/2010PASP..122..499E}{2010}, PASP, 122, 499
\bibitem[\protect\citeauthoryear{Eales et al.}{2018}]{Eales2018}
{Eales}, S., {Smith}, D., {Bourne}, N., {Loveday}, J., {Rowlands}, K., et al., \href{http://adsabs.harvard.edu/abs/2018MNRAS.473.3507E}{2018}, MNRAS, 473, 3507
\bibitem[\protect\citeauthoryear{Edge et al.}{2013}]{Edge2013}
Edge, A., Sutherland, W., Kuijken, K., Driver, S., McMahon, R., Eales, S., Emerson, J. P., \href{http://adsabs.harvard.edu/abs/2013Msngr.154...32E}{2013}, MNRAS, 409, 109
\bibitem[\protect\citeauthoryear{Eisenstein \& Hu}{1999}]{EisensteinHu1999}
Eisenstein D. J., Hu W., \href{http://adsabs.harvard.edu/abs/1999ApJ...511....5E}{1999}, ApJ, 511, 5
\bibitem[\protect\citeauthoryear{Foreman-Mackey et al.}{2013}]{Foreman-Mackey2013}
{Foreman-Mackey}, D. and {Hogg}, D.~W. and {Lang}, D. and {Goodman}, J. \href{http://adsabs.harvard.edu/abs/2013PASP..125..306F}{2013}, PASP, 125, 306
\bibitem[\protect\citeauthoryear{Frayer et al.}{2011}]{Frayer2011}
{Frayer}, D.~T., {Harris}, A.~I., {Baker}, A.~J., {Ivison}, R.~J., {Smail}, I., {Negrello}, M., et al. \href{http://adsabs.harvard.edu/abs/2011ApJ...726L..22F}{2011},  ApJ, 726, 22
\bibitem[\protect\citeauthoryear{Furlanetto et al.}{2018}]{Furlanetto2018}
 {Furlanetto}, C., {Dye}, S., {Bourne}, N., {Maddox}, S., {Dunne}, L., et al. \href{http://adsabs.harvard.edu/abs/2018MNRAS.tmp..149F}{2018},  MNRAS, 476, 961
 \bibitem[\protect\citeauthoryear{Geach et al.}{2012}]{Geach2012}
{Geach}, J.~E., {Sobral}, D., {Hickox}, R.~C., {Wake}, D.~A., {Smail}, I., et al., \href{http://adsabs.harvard.edu/abs/2012MNRAS.426..679G}{2012}, MNRAS, 426, 679
\bibitem[\protect\citeauthoryear{Geach et al.}{2013}]{Geach2013}
{Geach}, J.~E., {Chapin}, E.~L., {Coppin}, K.~E.~K., {Dunlop}, J.~S., {Halpern}, M., {Smail}, I., et al., \href{http://adsabs.harvard.edu/abs/2013MNRAS.432...53G}{2013},
\bibitem[\protect\citeauthoryear{George et al.}{2013}]{George2013}
{George}, R.~D., {Ivison}, R.~J., {Hopwood}, R., {Riechers}, D.~A., {Bussmann}, R.~S., {Cox}, P., et al., \href{http://adsabs.harvard.edu/abs/2013MNRAS.436L..99G}{2013}, MNRAS, 436, 99
\bibitem[\protect\citeauthoryear{Gonzalez-Nuevo et al.}{2005}]{GonzalezNuevo2005}
{Gonz{\'a}lez-Nuevo}, J., Toffolatti L., Argueso F., \href{http://adsabs.harvard.edu/abs/2005ApJ...621....1G}{2005}, ApJ, 621, 1
\bibitem[\protect\citeauthoryear{Gonzalez-Nuevo et al.}{2014}]{GonzalezNuevo2014}
{Gonz{\'a}lez-Nuevo}, J., {Lapi}, A., {Negrello}, M., {Danese}, L., {De Zotti}, et al., \href{http://adsabs.harvard.edu/abs/2014MNRAS.442.2680G}{2014}, MNRAS, 442, 2680
\bibitem[\protect\citeauthoryear{Gonzalez-Nuevo et al.}{2017}]{GonzalezNuevo2017}
{Gonz{\'a}lez-Nuevo}, J., {Lapi}, A., {Bonavera}, L., {Danese}, L., {de Zotti}, G., {Negrello}, et al., \href{http://adsabs.harvard.edu/abs/2017JCAP...10..024G}{2017}, JCAP, 10, 024
\bibitem[\protect\citeauthoryear{Granato et al.}{2004}]{Granato2004}
{Granato}, G.~L., {De Zotti}, G., {Silva}, L., {Bressan}, A., {Danese}, L., \href{http://adsabs.harvard.edu/abs/2004ApJ...600..580G}{2004}, ApJ, 600, 580
\bibitem[\protect\citeauthoryear{Griffin et al.}{2010}]{Griffin2010}
Griffin M. J., et al., \href{http://adsabs.harvard.edu/abs/2010A\%26A...518L...3G}{2010}, A\&A, 518, L3
\bibitem[\protect\citeauthoryear{Harris et al.}{2012}]{Harris2012}
{Harris}, A.~I., {Baker}, A.~J., {Frayer}, D.~T., {Smail}, I., {Swinbank}, A.~M., {Riechers}, D.~A., et al., \href{http://adsabs.harvard.edu/abs/2012ApJ...752..152H}{2012}, ApJ, 752, 152
\bibitem[\protect\citeauthoryear{Heath}{1977}]{Heath1977}
Heath D. J., \href{http://adsabs.harvard.edu/abs/1977MNRAS.179..351H}{1977}, MNRAS, 179, 351
\bibitem[\protect\citeauthoryear{Hickox et al.}{2012}]{Hickox2012}
Hickox R. C., Wardlow J. L., Smail I., Myers A. D., Alexander D. M., et al., \href{http://adsabs.harvard.edu/abs/2012MNRAS.421..284H}{2012}, MNRAS, 421, 284
\bibitem[\protect\citeauthoryear{Ivison et al.}{2011}]{Ivison2011}
Ivison R. J., Papadopoulos P. P., Smail I., Greve T. R., et al., \href{http://adsabs.harvard.edu/abs/2011MNRAS.412.1913I}{2011}, MNRAS, 412, 1913 
\bibitem[\protect\citeauthoryear{Ivison et al.}{2016}]{Ivison2016}
{Ivison}, R.~J., {Lewis}, A.~J.~R., {Weiss}, A., {Arumugam}, V., {Simpson}, J.~M., {Holland}, et al., \href{http://adsabs.harvard.edu/abs/2016ApJ...832...78I}{2016}, ApJ, 832, 78I
\bibitem[\protect\citeauthoryear{Jin et al.}{2018}]{Jin2018}
{Jin}, S., {Daddi}, E., {Liu}, D., {Smol{\v c}i{\'c}}, V., {Schinnerer}, E., et al., \href{http://adsabs.harvard.edu/abs/2018ApJ...864...56J}{2018}, ApJ, 864, 56
\bibitem[\protect\citeauthoryear{Lapi}{2011}]{Lapi2011}
{Lapi}, A., {Gonz{\'a}lez-Nuevo}, J., {Fan}, L., {Bressan}, A., {De Zotti}, G., {Danese}, L., et al., \href{http://adsabs.harvard.edu/abs/2011ApJ...742...24L}{2011}, ApJ, 742, 24
\bibitem[\protect\citeauthoryear{Landy \& Szalay}{1993}]{LandySzalay1993}
Landy S. D. \& Szalay A. S. \href{http://adsabs.harvard.edu/abs/1993ApJ...412...64L}{1993}, ApJ, 412, 64
\bibitem[\protect\citeauthoryear{Lawrence et al.}{2007}]{Lawrence2007}
Lawrence A., Warren, S. J., Almaini, O., Edge, A. C., Hambly N. C., et al., \href{http://adsabs.harvard.edu/abs/2007MNRAS.379.1599L}{2007}, , MNRAS, 379, 1599
\bibitem[\protect\citeauthoryear{Lewis, Challinor \& Lasenby}{2002}]{LewisChallinorLasenby2002}
Lewis A., Challinor A., Lasenby A., \href{http://adsabs.harvard.edu/abs/2000ApJ...538..473L}{2000}, Phys. Rev. D, 538, 473
\bibitem[\protect\citeauthoryear{Limber}{1954}]{Limber1954}
Limber D. N., \href{http://adsabs.harvard.edu/abs/1954ApJ...119..655L}{1954}, ApJ, 119, 655
\bibitem[\protect\citeauthoryear{Lupu et al.}{2012}]{Lupu2012}
{Lupu}, R.~E., {Scott}, K.~S., {Aguirre}, J.~E., {Aretxaga}, I., {Auld}, R., {Barton}, E., et al., \href{http://adsabs.harvard.edu/abs/2012ApJ...757..135L}{2012}, ApJ, 757, 135
\bibitem[\protect\citeauthoryear{Maddox et al.}{2010}]{Maddox2010}
Maddox S. J., Dunne L., Rigby E., Eales S., Cooray A., et al., \href{http://adsabs.harvard.edu/abs/2010A\%26A...518L..11M}{2010}, MNRAS, 518, 11
\bibitem[\protect\citeauthoryear{Maddox et al.}{2018}]{Maddox2018}
{Maddox}, S.~J., {Valiante}, E., {Cigan}, P., {Dunne}, L., {Eales}, S., et al., \href{http://adsabs.harvard.edu/abs/2018ApJS..236...30M}{2018}, ApJS, 236, 30
\bibitem[\protect\citeauthoryear{Magliocchetti et al.}{2014}]{Magliocchetti2014}
Magliocchetti M., Lapi A., Negrello M., De Zotti G., Danese L., \href{http://adsabs.harvard.edu/abs/2014MNRAS.437.2263M}{2014}, MNRAS, 437, 2263
\bibitem[\protect\citeauthoryear{Martin et al.}{2005}]{Martin2005}
{Martin}, D.~C., {Fanson}, J., {Schiminovich}, D., {Morrissey}, P., {Friedman}, P.~G., et al., \href{http://adsabs.harvard.edu/abs/2005ApJ...619L...1M}{2005}, ApJ, 619, 1
\bibitem[\protect\citeauthoryear{Mitchell-Wynne et al.}{2012}]{Mitchell-Wynne2012}
Mitchell-Wynne K., Cooray A., Gong Y., Bethermin M., Bock J., et al., \href{http://adsabs.harvard.edu/abs/2012ApJ...753...23M}{2012}, ApJ, 753, 23
\bibitem[\protect\citeauthoryear{Mo \& White}{2002}]{MoWhite2002}
{Mo}, H.~J., {White}, S.~D.~M., \href{http://adsabs.harvard.edu/abs/2002MNRAS.336..112M}{2002}, MNRAS, 336, 112
\bibitem[\protect\citeauthoryear{Murray et al.}{2013}]{Murray2013}
Murray S. G., Power C., Robotham A. S. G., \href{http://adsabs.harvard.edu/abs/2013A\%26C.....3...23M}{2013}, Astronomy and Computing, 3, 23
\bibitem[\protect\citeauthoryear{Nakamura \& Suto}{1997}]{NakamuraSuto1997}
Nakamura, T. T., \& Suto Y., \href{http://adsabs.harvard.edu/abs/1997PThPh..97...49N}{1997}, PThPh, 97, 49
\bibitem[\protect\citeauthoryear{Navarro, Frenk \& White}{1996}]{NFW1996}
Navarro J. F., Frenk C. S., White S. D. M., \href{http://adsabs.harvard.edu/abs/1996ApJ...462..563N}{1996}, ApJ, 462, 563
\bibitem[\protect\citeauthoryear{Navarro, Frenk \& White}{1997}]{NFW1997}
Navarro J. F., Frenk C. S., White S. D. M., \href{http://adsabs.harvard.edu/abs/1997ApJ...490..493N}{1997}, ApJ, 490, 493
\bibitem[\protect\citeauthoryear{Negrello et al.}{2010}]{Negrello2010}
{Negrello}, M., {Hopwood}, R., {De Zotti}, G., {Cooray}, A., {Verma}, A., {Bock}, J., {Frayer}, D.~T., et al., \href{http://adsabs.harvard.edu/abs/2010Sci...330..800N}{2010}, Science, 330, 800
\bibitem[\protect\citeauthoryear{Norberg et al.}{2009}]{Norberg2009}
Norberg P., Baugh C., M., Gaztanaga E. Croton D. J., \href{http://adsabs.harvard.edu/abs/2009MNRAS.396...19N}{2009}, MNRAS, 396, 19 
\bibitem[\protect\citeauthoryear{Oliver et al.}{2010}]{Oliver2010}
{Oliver}, S.~J., {Wang}, L., {Smith}, A.~J., {Altieri}, B., {Amblard}, A., et al., \href{http://adsabs.harvard.edu/abs/2010A\%26A...518L..21O}{2010}, A\&A, 518, 21
\bibitem[\protect\citeauthoryear{Pearson et al.}{2013}]{Pearson2013}
Pearson E. A., Eales S., Dunne L., Gonzalez-Nuevo J., Maddox S., et al., \href{http://adsabs.harvard.edu/abs/2013MNRAS.435.2753P}{2013}, MNRAS, 435, 2753
\bibitem[\protect\citeauthoryear{Peebles}{1980}]{Peebles1980}
{Peebles}, P.~J.~E., \href{http://adsabs.harvard.edu/abs/1980lssu.book.....P}{1980}, The Large Scale Structure of the Universe (Princeton: Princeton University Press)
\bibitem[\protect\citeauthoryear{Pilbratt et al.}{2010}]{Pilbratt2010}
Pilbratt G. L., Riedinger J. R., Passvogel T., Crone G., Doyle D., et al., \href{http://adsabs.harvard.edu/abs/2010A\%26A...518L...1P}{2010}, A\&A, 518, L1
\bibitem[\protect\citeauthoryear{{Planck Collaboration}}{2011}]{PlanckCollaboration2011_XVIII}
{Planck Collaboration}, et al., \href{http://adsabs.harvard.edu/abs/2011A\%26A...536A..18P}{2011}, A\&A, 536, A18
\bibitem[\protect\citeauthoryear{{Planck Collaboration XXX}}{2014}]{PlanckCollaboration2014_XXX}
{Planck Collaboration}, et al., \href{http://adsabs.harvard.edu/abs/2014A\%26A...571A..30P}{2014}, A\&A, 571, 30
\bibitem[\protect\citeauthoryear{Planck Collaboration XIII}{2016}]{PlanckCollaboration2016}
Planck Collaboration, et al., \href{http://adsabs.harvard.edu/abs/2016A\%26A...594A..13P}{2015}, A\&A, 594, 63
\bibitem[\protect\citeauthoryear{Poglitsch et al.}{2010}]{Poglitsch2010}
Poglitsch A.,  Waelkens C., Geis N., Feuchtgruber H., Vandenbussche B., et al., \href{http://adsabs.harvard.edu/abs/2010A\%26A...518L...2P}{2010}, A\&A, 518, L2
\bibitem[\protect\citeauthoryear{Roche \& Eales}{1999}]{RocheEales1999}
Roche N., Eales S. A., \href{http://adsabs.harvard.edu/abs/1999MNRAS.307..703R}{1999}, MNRAS, 307, 703
\bibitem[\protect\citeauthoryear{Scott et al.}{2006}]{Scott2006}
Scott S. E., Dunlop J. S., Serjeant S., \href{http://adsabs.harvard.edu/abs/2006MNRAS.370.1057S}{2006}, MNRAS, 370, 1057
\bibitem[\protect\citeauthoryear{Simpson et al.}{2014}]{Simpson2014}
Simpson J. M., Swinbank A. M., Smail I., et al., \href{http://adsabs.harvard.edu/abs/2014ApJ...788..125S}{2014}, ApJ, 788, 125
\bibitem[\protect\citeauthoryear{Smail et al.}{1997}]{Smail1997}
Smail I., Ivison R. J., \& Blain A. W., \href{http://adsabs.harvard.edu/abs/1997ApJ...490L...5S}{1997}, ApJL, 490, L5
\bibitem[\protect\citeauthoryear{Smith et al.}{2017}]{Smith2017}
{Smith}, M.~W.~L., {Ibar}, E., {Maddox}, S.~J., {Valiante}, E., {Dunne}, L., {Eales}, S., et al., \href{http://adsabs.harvard.edu/abs/2017ApJS..233...26S}{2017}, arXiv:1712.02361
\bibitem[\protect\citeauthoryear{Swinbank et al.}{2014}]{Swinbank2014}
Swinbank A. M., Simpson J. M., Smail I., Harrison C. M., Hodge J. A., et al., \href{http://adsabs.harvard.edu/abs/2014MNRAS.438.1267S}{2014}, MNRAS, 438, 1267
\bibitem[\protect\citeauthoryear{Tinker et al.}{2010}]{Tinker2010}
{Tinker}, J.~L., {Robertson}, B.~E., {Kravtsov}, A.~V., {Klypin}, A., et al., \href{http://adsabs.harvard.edu/abs/2010ApJ...724..878T}{2010}, ApJ, 724, 878
\bibitem[\protect\citeauthoryear{Valiante et al.}{2016}]{Valiante2016}
Valiante E., Smith M. W. L., Eales S., Maddox S. J., et al., \href{http://adsabs.harvard.edu/abs/2016MNRAS.462.3146V}{2016}, MNRAS, 462, 3146
\bibitem[\protect\citeauthoryear{van Kampen et al.}{2012}]{vanKampen2012}
van Kampen E., Smith D. J. B., Maddox S., Hopkins A. M., et al., \href{http://adsabs.harvard.edu/abs/2012MNRAS.426.3455V}{2012}, MNRAS, 426, 3455
\bibitem[\protect\citeauthoryear{Viero et al.}{2009}]{Viero2009}
Viero M. P., Ade P. A. R., Bock J. J., et al., \href{http://adsabs.harvard.edu/abs/2009ApJ...707.1766V}{2009}, ApJ, 707, 1766
\bibitem[\protect\citeauthoryear{Viero et al.}{2013}]{Viero2013}
{Viero}, M.~P., {Wang}, L., {Zemcov}, M., {Addison}, G., {Amblard}, A., {Arumugam}, V., et al., \href{http://adsabs.harvard.edu/abs/2013ApJ...772...77V}{2013}, ApJ, 772, 77
\bibitem[\protect\citeauthoryear{Waddington et al.}{2007}]{Waddington2007}
Waddington I., Oliver S. J., Babbedge T. S. R., Fang F., Farrah D., et al., \href{http://adsabs.harvard.edu/abs/2007MNRAS.381.1437W}{2007}, MNRAS, 381, 1437
\bibitem[\protect\citeauthoryear{Webb et al.}{2003}]{Webb2003}
{Webb}, T.~M., {Eales}, S., {Foucaud}, S., {Lilly}, S.~J., {McCracken}, H., {Adelberger}, K., et al., \href{http://adsabs.harvard.edu/abs/2003ApJ...582....6W}{2003}, ApJ, 582, 6 
\bibitem[\protect\citeauthoryear{{Wei{\ss}} et al.}{2009}]{Weiss2009}
{Wei{\ss}}, A., {Kov{\'a}cs}, A., {Coppin}, K., {Greve}, T.~R., {Walter}, F., {Smail}, I., et al., \href{http://adsabs.harvard.edu/abs/2009ApJ...707.1201W}{2009}, ApJ, 707, 1201
\bibitem[\protect\citeauthoryear{Wilkinson et al.}{2017}]{Wilkinson2017}
Wilkinson A., Almaini O., Chen C., Smail. I., et al., \href{http://adsabs.harvard.edu/abs/2017MNRAS.464.1380W}{2017}, MNRAS, 464, 1380
\bibitem[\protect\citeauthoryear{Williams et al.}{2011}]{Williams2011}
{Williams}, C.~C., {Giavalisco}, M., {Porciani}, C., {Yun}, M.~S., {Pope}, A., et al., \href{http://adsabs.harvard.edu/abs/2011ApJ...733...92W}{2011}, ApJ, 733, 92
\bibitem[\protect\citeauthoryear{Wright et al.}{2010}]{Wright2010}
{Wright}, E.~L., {Eisenhardt}, P.~R.~M., {Mainzer}, A.~K., {Ressler}, M.~E., {Cutri}, R.~M., et al., \href{http://adsabs.harvard.edu/abs/2010AJ....140.1868W}{2010}, AJ, 140, 1868
\bibitem[\protect\citeauthoryear{Zehavi et al.}{2011}]{Zehavi2011}
{Zehavi}, I., {Zheng}, Z., {Weinberg}, D.~H., {Blanton}, M.~R., {Bahcall}, N.~A., et al., \href{http://adsabs.harvard.edu/abs/2011ApJ...736...59Z}{2011}, ApJ, 736, 59
\bibitem[\protect\citeauthoryear{Zheng et al.}{2005}]{Zheng2005}
Zheng Z., Berlind A. A., Weinberg D. H., Benson A. J., Baugh C. M., et al., \href{http://adsabs.harvard.edu/abs/2005ApJ...633..791Z}{2005}, ApJ, 633, 791
\end{thebibliography}
\end{document}